\documentclass[pre,aps,twocolumn,superscriptaddress,showpacs]{revtex4}

\usepackage[dvips]{graphicx}
\usepackage{amssymb,amsfonts,amsmath}
\usepackage{color}
\usepackage{ulem}

\newcommand{\rv}{{\mathbf r}}

\newcommand{\J}{{\bf J}}

\newcommand{\F}{{\bf F}}

\newcommand{\kT}{k_{\rm B}T}

\begin{document}

\title{Phase separation on the sphere: Patchy particles and self-assembly}

\author{M.C.~Bott}

\author{J.M.~Brader}
\address{Soft Matter Theory, 
  University of Fribourg, CH-1700 Fribourg, Switzerland}

\begin{abstract}
Motivated by observations of heterogeneous domain structure on the surface of cells,  
we consider a minimal model to describe the dynamics of phase separation on the surface of a 
spherical particle. 
Finite size effects on the curved particle surface lead to the formation of long-lived,  
metastable states for which the density is distributed in patches over the particle surface. 
We study the time evolution and stability of these states as a function of both the particle 
size and the thermodynamic parameters. 
Finally, by connecting our findings with studies of patchy particles we consider the 
implications for self-assembly in many-particle systems.  
\end{abstract}


\maketitle


\section{Introduction}

Phase separation in bulk systems can proceed via a number of distinct physical 
mechanisms (spinodal decomposition, heterogeneous or homogeneous nucleation) and 
is generally well understood. 
However, the dynamical processes involved are less clear when the system is subject 
to some form of spatial confinement. 
This confinement can arise from the presence of external fields, representing, for example, 
substrates or random obstacles, but can also be imposed by the geometry of the embedding 
space. The latter type of confinement is particularly relevant to biological cells, 
for which the mobile fluid particles constituting the cell membrane are constrained to 
lie on the surface of a (roughly) spherical body. 
These membranes exhibit stable domains, the spatial distribution of which is 
important for many cell properties, e.g.~adhesion 
\cite{lenne2009,lingwood2010}. The composition of these domains and their distribution 
over the surface of the cell dictates to a large extent the 
interaction forces between different cells and thus the self-assembly behavior 
of many-cell systems.

One common view is that the domains on the surface of a cell are a consequence of 
an arrested or incomplete phase 
separation, however, it remains to be established whether the observed 
states are permanent or metastable in character. 
If the hetrogeneous domain structure on the cell surface is an equilibrium state, 
then some stabilizing mechanism is required;~the line-tension incurred by interfaces between 
domains would make inhomogeneous phases energetically unfavorable when 
compared to a fully phase separated system.
Computer simulation studies of simple model systems \cite{vink1,vink2} have shown 
that the size, composition and dynamics of membrane domains can be regulated by introducing 
randomly located, immobile objects. 
These obstacles, embedded within the two-dimensional fluid, serve to hinder
macroscopic phase separation and act as a source of quenched disorder. 
It has been proposed in Refs. \cite{vink1,vink2} that the quenched disorder found in real 
cells, provided by fixed cytoskeletal proteins, could be the key stabilizing mechanism.  

An alternative scenario is that the domains are long-lived metastable, rather than equilibrium, 
states. It is well-known from studies of phase separation \cite{Li1994,Lee1994} in bulk that quenching the 
thermodynamic parameters to a statepoint close to the spinodal will result in very slow phase 
separation dynamics. 
Following the quench, spherical domains of the minority phase form and then slowly merge together, 
a process known as Ostwald ripening \cite{onuki}.
The ripening process could possibly be slowed down, or even 
arrested entirely, by the presence of small quantities of an additional species, 
which sits preferentially at the interfaces between domains.

A requirement for studying domain formation is an understanding of diffusion processes 
on the sphere.  
Such studies can be found in the literature in a variety of contexts. 
The diffusion of non-interacting particles on a spherical surface has been addressed using 
analytical methods by Ghosh {\it et al.} \cite{Ghosh2012}. 
Marenduzzo and Orlandini have used numerical methods to study diffusive motion on general 
curved surfaces and investigated the coupling between phase separation and local curvature 
\cite{Marenduzzo2013}. 
Fischer and Vink performed many-body simulations on a spherical surface, with the aim to 
optimize the boundary conditions for simulations of first-order transitions in finite-size systems 
\cite{Fischer2010}. 

Going beyond single cell properties, 
assemblies of spherical cells exhibit nontrivial 
interactions, both with each other and with external substrates. 
The interaction potential between a pair of cells is strongly influenced 
by the distribution and size of the domains covering its surface. 
In this sense, cells may be regarded as a naturally occurring type of 
`patchy particle'; the term given to particles with distinct surface sites generating 
anisotropic interparticle interactions. 
While synthetically fabricated patchy particles have attractive 
interaction patches strategically arranged on their surface \cite{likos_review}, the 
domains covering the cell emerge as a result of self-organization. 
When multiple cells are present in a crowded environment the influence of competing physical 
mechanisms, acting both within each cell membrane (line tension, quenched disorder) and between 
different cells, can generate a complex domain structure.  

The phase behavior and equilibrium microstructure of synthetic patchy particles depends 
upon the number, spatial distribution and attraction strength of the interaction patches. 
For example, spherical particles with just two attractive patches will tend to form 
polymer-like chains, wheras three-patch particles will assemble into open gel-like 
structures (`empty' liquids) 
\cite{bianchi}. 
Recent developments in the controlled fabrication of patchy particles have raised hopes that 
materials with desired properties may be tailored by prescribing the number and geometrical 
arrangement of the patches \cite{glotzer,likos_review}. 
In order to understand the collective behavior of natural patchy particles, for which the 
domains self-organize, it is necessary to understand first the dynamical processes occurring 
on the surface of individual cells. 

In this paper we investigate how phase separation on the surface of a spherical 
body can give rise to different domain structures, and then infer how these domains could influence the 
self-assembly in systems consisting of many spherical bodies. 
We do not seek to describe real biological cells, but rather take these 
as a motivation for the construction of simple models capturing generic physical features. 
We will focus first on single particle properties, investigating how the domains form on the 
particle surface under various conditions, before proceeding to study how these 
domains may influence interparticle interactions.  
In section \ref{theory} we outline the model system to be considered, the theoretical method employed and 
the numerical methods used to solve our equations.  
In section \ref{results} we investigate the domain formation on a single spherical body 
and infer from this the likely consequences for many-body self-assembly. 
Finally, in section \ref{conclusions} we discuss our findings and provide an outlook.


\section{Theory}\label{theory}

We will investigate the demixing of a binary fluid on the two-dimensional surface 
of a large spherical particle. 
In order to avoid any confusion with terminology, we will 
henceforth refer to the large particle as the `meso-particle' and the smaller, 
mobile particles constituting the fluid on its surface as the `surface particles'. 
As we are interested in the phenomenology of phase separation and domain formation 
we choose for convenience a very simple microscopic model, the Gaussian core model 
(GCM), to represent the surface particles. In the present study the GCM is employed 
simply because of its generic demixing properties, rather than as an approximation to any 
specific physical system.  
The collective behavior of the GCM on the meso-particle surface is treated using 
a well-established mean-field density functional theory. 

\subsection{The Gaussian core model}

To represent the surface particles, we consider a model binary mixture in which the 
particles interact via the soft repulsive pair potential 
\begin{equation}
\label{eq:gaussian_interaction}
 \beta v_{ij}(r) = \beta \epsilon_{ij} \exp\{ -r^2 / R_{ij}^2\},
\end{equation}
where $\beta\!=\!(k_BT)^{-1}$ and the non-negative parameters $\epsilon_{ij}$ and $R_{ij}$ 
determine the strength and range, 
respectively, of the interaction between species $i$ and $j$. 
The GCM was introduced by Stillinger \cite{Stillinger1976} 
to study phase separation in binary mixtures and has since been 
studied intensively, both in bulk and at interfaces 
\cite{Archer2001, Archer2002, Archer2003, Archer2004, Archer2005, Archer2006}. 
The model has the advantage that a simple mean-field approximation to the free energy provides 
good agreement with computer simulation data \cite{Louis2000}. 

When calculating the interaction between surface particles the separation $r$ entering the pair 
potential \eqref{eq:gaussian_interaction} is taken to be the direct, straight-line 
distance (cutting through the meso-sphere), rather than the length of the arc around the 
surface of the meso-sphere.

\subsection{Mean-field free energy functional}

To describe the collective behaviour of the surface particles 
we use an approximation to the two-dimensional Helmholtz free energy functional
\begin{align}
\label{eq:helmolzfreenergy}
 \beta \mathcal{F}[\{\rho_q(\rv,t)\}] 
 &= \sum\limits_{q} \int d\mathbf{r} \rho_{q}(\mathbf{r}) \bigl (\ln(\rho_{q}(\mathbf{r})) - 1 \bigr ) \\
 &\!\!\!
+ \frac{1}{2} \sum\limits_{ql} \int d\mathbf{r}\! \int d\mathbf{r'} \rho_{q}(\mathbf{r})\rho_{l}(\mathbf{r'}) \beta v_{ql}(|\mathbf{r - \mathbf{r'}}|) \nonumber, 
\end{align}
where the first and second terms provide the ideal and excess (over ideal) contributions, respectively. 
The subscripts $q$ and $l$ are species labels and the notation $[\{\rho_q(\rv,t)\}]$ indicates a functional 
dependence on the one-body density profiles of all species. 
We set the (physically irrelevant) thermal wavelength equal to unity. 
For a binary mixture the species indicies are restricted to the values $q,l=1,2$.  
In bulk, the number density of species $q$ is $\rho_q\!=\!N_q/V$, where $V$ is the area in the 2d case. The total density 
is $\rho=\rho_1 + \rho_2$.  

It is convenient to introduce a concentration variable $x = N_2/N$, which enables 
the species labeled densities to be expressed as $\rho_1 =(1-x)\rho$ and $\rho_2 =x \rho$.
In these variables the bulk free energy per particle 
consists of a sum of two terms, $f\equiv F/N=f_{\rm id} + f_{\rm ex}$. 
The ideal part is given by 
%
%
%
\begin{eqnarray}\label{ideal_perparticle}
\beta f_{\rm id} 
&=& \ln(\rho) - 1 + (1-x) \ln(1-x) + x\ln(x), 
\end{eqnarray}
%
where the contribution $(1-x)\ln(1-x) + x\ln(x)$ is due to the entropy of mixing.
The reduced bulk excess free energy per particle is given by
\begin{equation}
 \beta f_{\rm ex} = \frac{1}{2\rho} \big(\rho_1 \rho_1 \hat{v}_{11}(0) + 2\rho_1 \rho_2 \hat{v}_{12}(0) + \rho_2 \rho_2 \hat{v}_{22}(0) \big).
\end{equation}
where $\hat{v}_{ij}(0)$ is the Fourier transform of the pair potential at zero wavevector 
$\hat{v}_{ij}(k=0)\!=\!\epsilon_{ij}^* R_{ij}^2 \pi$ and $\epsilon_{ij}^* = \beta \epsilon_{ij}$.
%
%

%
%
Expressing $\rho_1$ and $\rho_2$ in terms of the concentration variable $x$, one obtains
\begin{equation}\label{excess_perparticle}
\beta f_{\rm ex} = \frac{1}{2} \rho 
\left(
(1-x)^2 \hat{v}_{11}(0) + 2x(1-x) \hat{v}_{12}(0) + x^2\hat{v}_{22}(0)
\right).
\end{equation}
%
%
%
%
When the total density $\rho$ becomes sufficiently large the GCM demixes. 
To obtain the coexistence curve (binodal) both the chemical potential of each species and the pressure 
have to be set equal in the coexisting phases (see Appendix 1).


\subsection{Microscopic dynamics of surface particles}\label{smol_section}

If we assume that the momentum degrees of freedom of the surface particles 
equilibrate much faster than their positions, then the motion of the surface particles
may be modelled using Brownian dynamics \cite{gardiner}.
For a multi-component system the configurational probability density, $\Psi(\{\rv_{iq}\},t)$, describes the 
probability to find a given particle configuration at time $t$, where $\rv_{iq}$ is the coordinate 
of the $i$th particle of species $q$. 
Given an initial state, the time evolution of $\Psi(\{\rv_{iq}\},t)$ is given by the 
Smoluchowski equation \cite{dhont_book}
\begin{align}
  \frac{\partial \Psi(\{\rv_{iq}\},t)}{\partial t} = 
  -\sum_{i}\sum_{q} \frac{\partial}{\partial \rv_{iq}} \cdot\J_{iq}(\{\rv_{iq}\},t),
\label{EQsmol}
\end{align}
where the sums are taken over all particles and species. 
The current of particle $i$ of species $q$ is given by
\begin{align}\label{EQsmol_current} 
\J_{iq}(\{\rv_{iq}\},t)&= \gamma_{q}^{-1}\Psi(\{\rv_{iq}\},t) \big[ 
  \F_{iq}(\{\rv_{iq}\},t) 
\notag\\
&\hspace*{0.4cm}- \kT\frac{\partial}{\partial \rv_{iq}} \ln \Psi(\{\rv_{iq}\},t)\big]. 
\end{align} 
where  $\gamma_q=\kT/D_{\! q}$ is a friction coefficient, $D_{\!q}$ is the bare diffusion coefficient 
of species $q$ and $k_BT$ is the thermal energy. 
We will henceforth assume, for simplicity, that all species have equal friction coefficient, 
$\gamma_q=\gamma$. 
The total force, $\F_{iq}(\{\rv_{iq}\})$, is the sum of contributions from 
interactions and external fields.



%
\begin{figure}
\vspace*{-2cm}
  \hspace*{0.cm}\includegraphics[width=9cm]{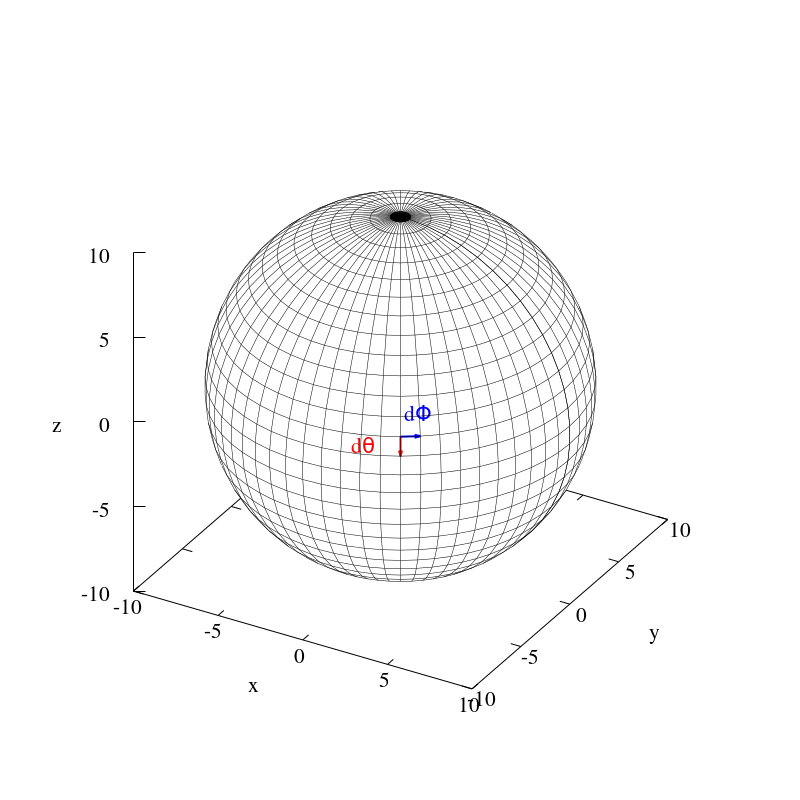}
\vspace*{-1.cm}\\
 \caption[numerical grid]{The numerical grid (here for $M=60$ and $N=30$) used to 
calculate density profiles on the meso-particle surface. 
 The angular increment $d\phi\!=\!2\pi/M$ points from west to east and the anglular increment 
$d\theta\!=\!pi/N$ points from the north pole to the south pole. 
This discretization leads to an oversampling of points around the two poles.}
 \label{fig:numerical_grid}
\end{figure}

\subsection{Dynamical density functional theory}\label{DDFT}

To study phase separation on the surface of a meso-particle we 
will focus on the dynamics of the one-body density of the surface particles. 
This can be obtained using 
dynamical density functional theory (DDFT) \cite{Archer2003,marconi1999}. Within this approach 
the time evolution of the density of species $q$ is given by a 
generalized diffusion equation
\begin{eqnarray}
\frac{\partial \rho_q(\rv,t)}{\partial t} = 
\frac{\partial}{\partial \rv}\cdot\Bigg[ \gamma^{-1} \rho_q(\rv,t) 
\frac{\partial}{\partial \rv}\frac{\delta 
\mathcal{F}[\{\rho_q(\rv,t)\}]}{\delta \rho_q(\rv,t)} \Bigg].
\label{stDDFT}
\end{eqnarray} 
The DDFT equation of motion \eqref{stDDFT} is obtained from the many-body Smoluchowski equation 
\eqref{EQsmol} by, (i) integrating over all but one of the particle coordinates, 
(ii) approximating the interaction forces using the equilibrium free energy functional. 
This second step constitutes an adiabatic assumption. As the adiabatic approximation is well 
documented we refer the interested reader to Refs.\cite{Archer2003} and 
\cite{reinhardtbrader} for a detailed derivation of equation \eqref{stDDFT}.


\subsection{Numerical implementation}\label{numerics}

To solve the DDFT equation of motion \eqref{stDDFT} on the surface of a meso-sphere we must 
define an appropriate numerical grid. 
The chosen grid should enable accurate finite difference schemes for 
calculating the gradient and divergence of scalar/vector fields, as well as an 
efficient method to compute the convolution of two scalar fields. 
We find that for the present application the most simple-minded approach is, in fact, 
the best choice: we parametrize the sphere 
using the spherical polar angles $\theta$ and $\phi$. 
In the following subsections we report relevant technical details of our numerical solution 
of \eqref{stDDFT}. 


\begin{figure}
\begin{center}
 \includegraphics[width=7.5cm]{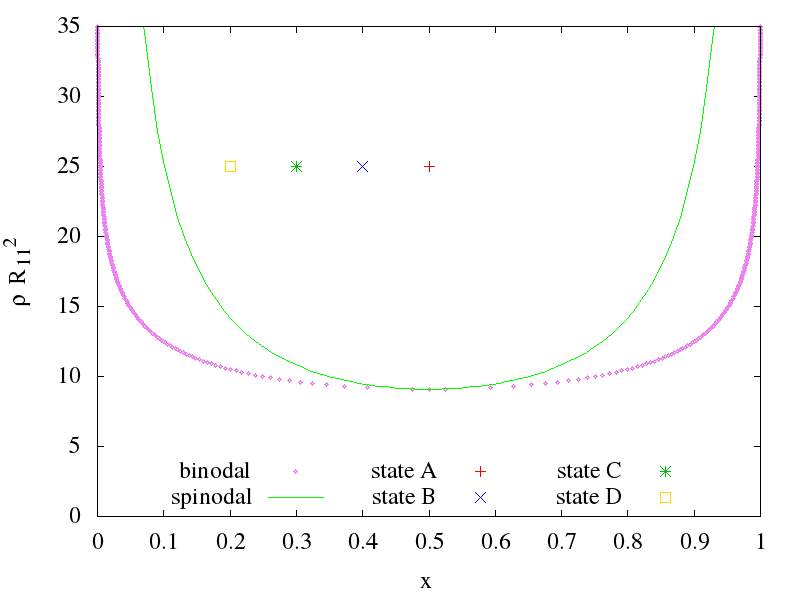}
 \caption{Phase diagram of the symmetric GCM for the surface particles in an 
infinite, two-dimensional, planar system. 
The parameters used are $R_{11}\!=\!R_{22}\!=\!R_{12}\!=\!1$, 
$\epsilon_{11}^*\!=\!\epsilon_{22}^*\!=\!2$ and 
$\epsilon_{12}^*\!=\!1.035\,\epsilon_{11}^*$. 
The critical point is located at $\rho R_{11}^2\!=\!9.094568$, $x\!=\!0.5$.}
\label{fig:binodal_volume}
\end{center}
\end{figure}	


{\it Numerical grid and finite differences:} 
We parameterize the surface of a meso-sphere of radius $R$ using the angles 
$\phi\!\in\![0,2\pi)$ and $\theta\!\in\![0,\pi]$.  
The $\phi$\,-range is divided into $M$ equally spaced points with spacing $d\phi\!=\!2 \pi/M$
and the $\theta$-range in $N$ points with spacing $d\theta\!=\!\pi/N$.
To avoid the singularity at the north ($\theta\!=\!0$) and south 
($\theta\!=\!\pi$) poles we exclude these two points and start our $\theta$ grid at 
$\theta_0\!=\!d\theta/2$ and end it at $\theta_{N-1}\!=\!(N-1) d\theta + d\theta/2 = \pi - d\theta/2$.
From Fig.\ref{fig:numerical_grid} it is evident that the pole regions suffer from oversampling 
when compared to the area around the equator.
However, this disadvantage is compensated by the ease with 
which finite differences may be calculated.  
All fields can be stored in $M\!\times\!N$ arrays and neighboring entries in the array correspond to 
physical neighbors on the sphere. The only complication arises on the edges, 
$\theta_0\!=\!d\theta/2$ and $\theta_{N-1}\!=\!\pi\!-\!d\theta/2$. 
Details of our finite difference scheme are given in Appendix 2.





%
\begin{figure}
\begin{center}
\includegraphics[width=0.5\textwidth]{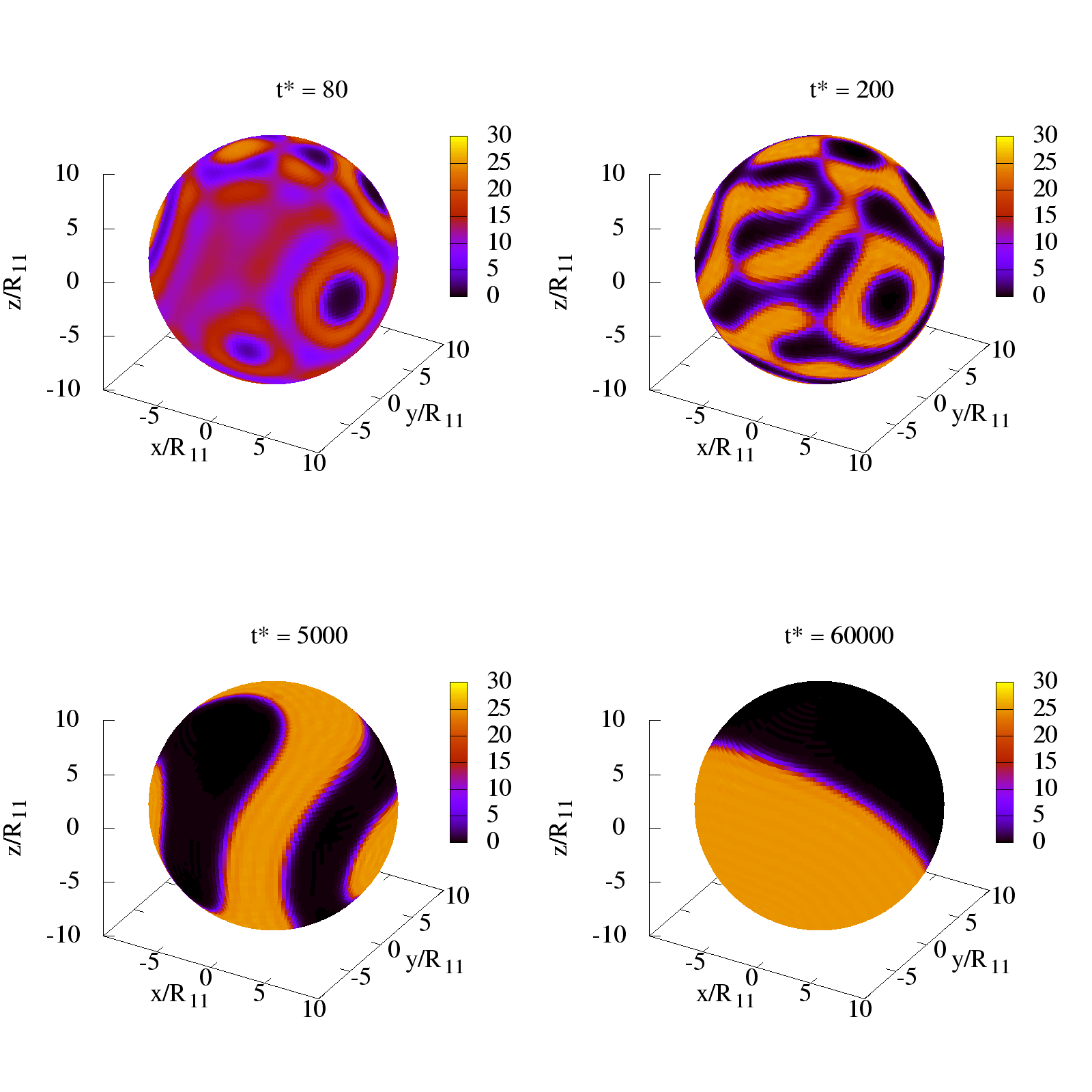}
\caption{{\bf Larger sphere with $\mathbf{x=0.5}$}.
The time-evolution of the density of species 1 with 
$\rho R_{11}^2 = 25$, confined on a meso-sphere of radius $R=10 R_{11}$.
The times shown are $t^*\!=\!tD/R_{11}^2 = 80, 200, 5000$ and $60000$. 
The process of spinodal decomposition leads to characteristic density inhomogeneities. 
In the long-time limit the line tension is minimized when the interface maps a great circle. 
%
}
\label{fig:spinodal_decomposition_x_0.5}
\end{center}
\end{figure}

\begin{figure}
\begin{center}
\includegraphics[width=0.5\textwidth]{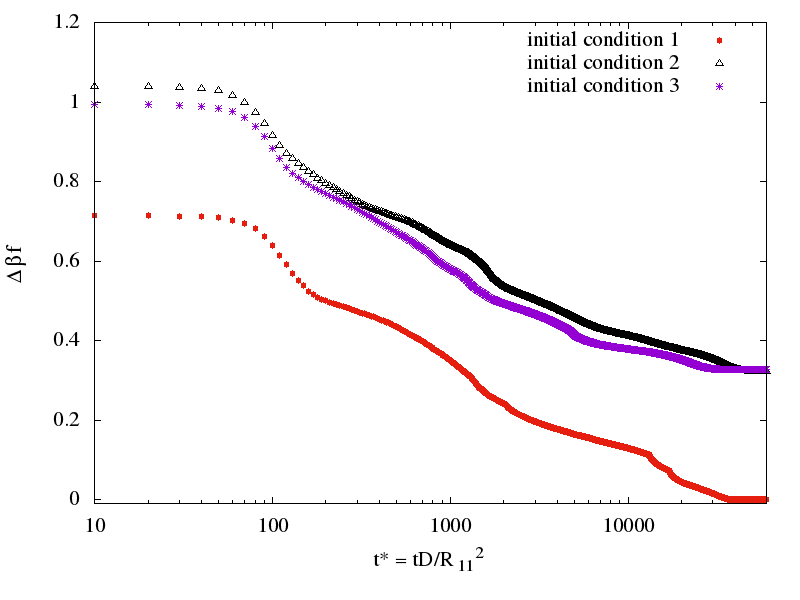}
\caption{{\bf Larger sphere with $\mathbf{x=0.5}$}.
Time-evolution of the free energy per particle for three different initial conditions of a system with 
$\rho R_{11}^2 = 25$. 
Initial condition 1 corresponds to the data shown in 
Fig.~\ref{fig:spinodal_decomposition_x_0.5} 
}
\label{fig:free_energy_comparison}
\end{center}
\end{figure}


{\it Convolutions:} The nonlocal approximation to the free energy, Eq.~\eqref{eq:helmolzfreenergy}, 
generates in Eq.~\eqref{stDDFT} convolution integrals of the form
\begin{equation}\label{convolution}
 \int d\Omega' f(\phi',\theta') \,g(\mathbf r - \mathbf{r'}),
\end{equation}
where $|\mathbf{r}|\!=\!|\mathbf{r'}|\!=\!R$ and both $f$ and $g$ are scalar functions.
Convolutions on the surface of a unit-sphere can be efficiently computed by expanding 
the scalar fields $f(\phi,\theta)$ and $g(\phi,\theta)$ in spherical harmonics
\begin{equation}\label{harmonic_expansion}
f(\phi,\theta) = \sum_{l=0}^{L} \sum_{|m| \le l} a_l^m Y_l^m(\phi,\theta).
\end{equation}
In principle an infinite number of terms are required, but in practice the series may be 
truncated at a finite value of $L$. 
It follows from orthogonality, $\int d\Omega \, \bar Y_{l'}^{m'}Y_{l}^m\!=\!\delta_{l',l}\delta_{m',m}$, 
that the coefficients are 
\begin{equation}
 a_l^m = \int d\Omega \, \bar Y_l^m(\phi,\theta)f(\phi,\theta).
\end{equation}
To compute the convolution \eqref{convolution} we make use of the fact that for two functions 
$f$ and $g$ defined on the unit sphere, the transform of the convolution is given by a 
pointwise product of the transforms, namely
\begin{equation}\label{SHconvolution}
 (f * g)_l^m = \sqrt{\frac{4\pi}{2l+1}} a_l^m \cdot  b_l^0,
\end{equation}
where $b_l^0 = \int d\Omega \, \bar Y_l^0(\phi,\theta)g(\phi,\theta)$. A proof of this statement and further insight on the method can be found in Ref.~\cite{Driscoll1994}. 
Extension of the convolution theorem \eqref{SHconvolution} to spheres of non-unit 
radius simply requires that equation \eqref{SHconvolution} be multiplied by a factor $R^{2}$. 
Spherical harmonic transforms were performed using an open source C library \cite{ccSHT,Frigo2005}. 
The computational effort for one transform is of order $(M\times N)^{3/2}$.



{\it Time Integration:} When solving \eqref{stDDFT} the spatial grid spacing
imposes a bound on the maximum stepsize $dt$ which can be used to calculate the time-evolution. 
Beyond a critical value of $dt$ the time-integration becomes unstable. 
This is the main drawback of our chosen spatial grid; the local oversampling 
around the poles leaves very little room to adjust the (global) stepsize $dt$. 
It is thus necessary to choose a value of $dt$ sufficiently small that the regions 
around the poles remain stable.  
The most reliable method to evolve 
\eqref{stDDFT} is simple Euler Integration. 
More sophisticated methods, such as Runge-Kutta integration combined with adaptive 
stepsize, do not lead to any significant increase in performance.


\section{Results}\label{results}


The surface of a meso-sphere of radius $R$ represents a finite size system and thus 
does not admit a true phase transition. 
Nevertheless, provided that a sufficient number of surface particles are present, then 
the phase diagram of an infinite 
planar system offers a useful guide when calculating density dynamics on the meso-sphere. 
The bulk phase diagram of an infinite planar system is shown in Fig.\ref{fig:binodal_volume} 
for the parameters $R_{11}\!=\!R_{22}\!=\!R_{12}\!=\!1$, 
$\epsilon_{11}^*\!=\!\epsilon_{22}^*\!=\!2$ 
and $\epsilon_{12}^*\!=\!1.035 \epsilon_{11}^*$. 
Statepoints at which we perform detailed calculations are indicated. 

\begin{figure}
\begin{center}
\includegraphics[width=0.5\textwidth]{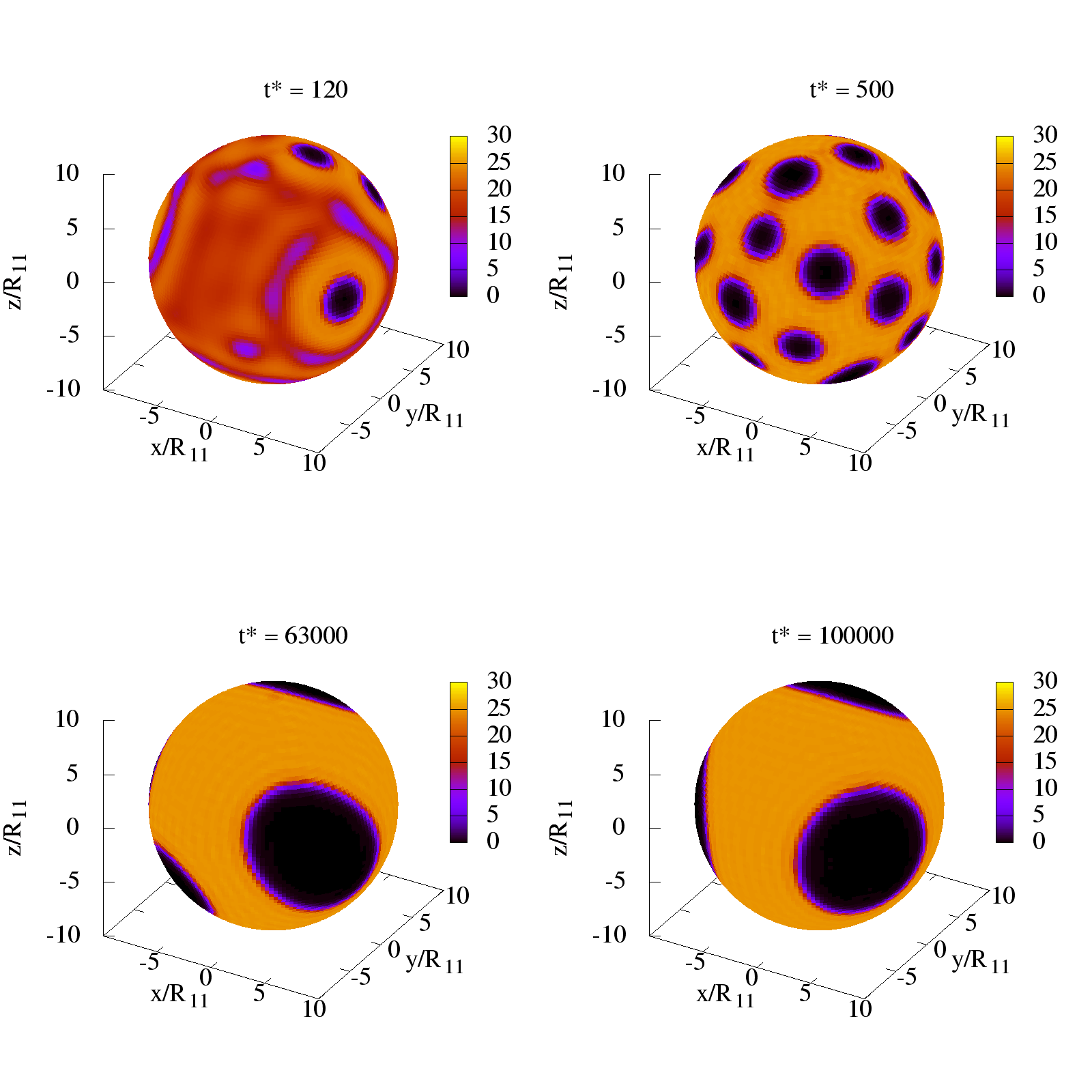}
\caption{{\bf Larger sphere with $\mathbf{x=0.3}$}.
-evolution of the density with $\rho R_{11}^2 = 25$ on a meso-particle with 
$R=10 R_{11}$.
The state at $t^*=10^5$ is not the final state, but rather a very long-lived metastable state with 
five patches (two of which are here located around the back of the meso-sphere). 	
}
\label{fig:spinodal_decomposition_x_0.3}
\end{center}
\end{figure}



We consider first a meso-sphere of radius $R=10 R_{11}$ with a total density of surface fluid 
particles $\rho R_{11}^2 = 25$ and composition $x=0.5$, corresponding to statepoint $A$ in the phase 
diagram (see Fig.~\ref{fig:binodal_volume}). 
%
%
%
%
In Figure \ref{fig:spinodal_decomposition_x_0.5} we show the density profile of species 1 at four 
different times. 
The initial condition is chosen by adding to a constant density several randomly located 
density peaks and dips of small amplitude.

\begin{figure}
\begin{center}
\includegraphics[width=0.5\textwidth]{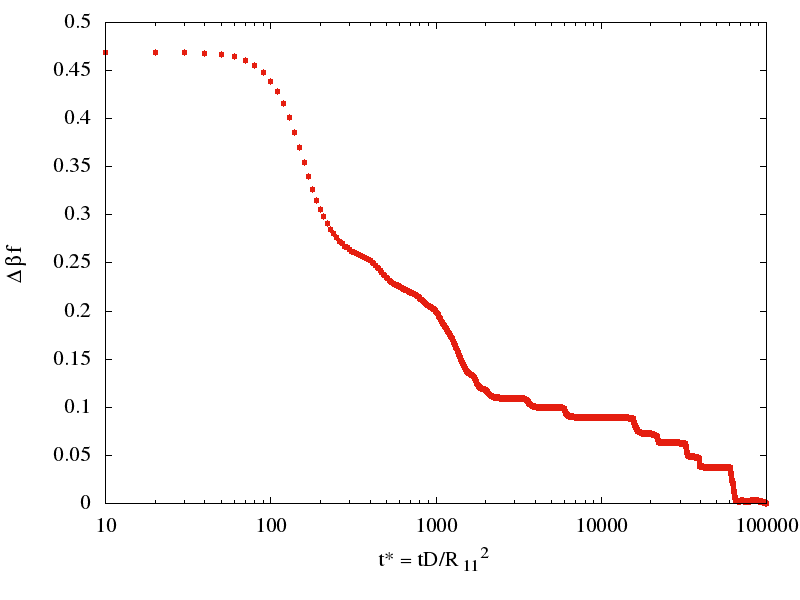}
\caption{
{\bf Larger sphere with $\mathbf{x=0.3}$}.
Time-evolution of the free energy at statepoint $\rho R_{11}^2 = 25$, corresponding to 
the density profiles shown in figure \ref{fig:spinodal_decomposition_x_0.3}. 
Sudden decreases in the free energy correspond to events where two domains merge. 
This contrasts with the smoother decay observed in 
Fig.~\ref{fig:free_energy_comparison}. 
}
\label{fig:free_energy_x_0.3}
\end{center}
\end{figure}

After a time $t^*\!=\!tD/R_{11}^2\!=\!80$ the spinodal instability becomes clearly visible on the scale 
of the figure. 
For later times (we show $t^*\!=\!200$ and $t^*\!=\!5000$) domains form and evolve as the system seeks 
to minimize the length of the boundary between the two phases. 
%
At the longest time for which we performed numerical calculations, $t^*\!=\!60000$, the interfacial region lies 
on a great circle, which is a consequence of the chosen composition $x\!=\!0.5$. 
We note that the orientation of the final phase-separated state is not correlated with the underlying numerical 
grid, thus suggesting that our chosen discretization does not introduce any artificial bias into 
the phase separation dynamics. 

As the DDFT is an adiabatic theory we can track the time evolution of the free energy. 
This is shown 
in Fig.~\ref{fig:free_energy_comparison} for three different initial conditions where we plot the free energy per particle $\beta f$ minus the long-time value of the free energy.   
%
Aside from slight differences arising from different initial conditions, the general behavior of the 
free energy relaxation is very similar for all cases investigated; a rapid initial relaxation is followed 
by a slow decay to equilibrium.  

We next consider a composition $x\!=\!0.3$, corresponding to 
statepoint C in Fig.\ref{fig:binodal_volume}. 
The density is shown in Fig.~\ref{fig:spinodal_decomposition_x_0.3} for four different times. 
In contrast to the behavior for $x\!=\!0.5$, the initial stage of the evolution for $x\!=\!0.3$ 
is characterized by the formation of circular islands of the minority phase which 
then slowly merge together; a process known as Ostwald ripening \cite{onuki}. 

In Fig.~\ref{fig:free_energy_x_0.3} we show the corresponding free energy as a function of time.
The free energy decreases rapidly whenever two circular patches merge, however, these merging events become 
less frequent as time progresses (note the logarithmic timescale). 
Even after $t^*=10^5$ the system has still not attained its final state, but the free energy 
shows no significant further decrease.  
The final state shown in Fig.~\ref{fig:free_energy_x_0.3} proved to be very stable; the expected completely 
phase separated state could not be obtained within the available computation time. 
For the duration that the surface fluid is trapped in this metastable state, which according to 
our calculations survives many tens of thousands of Brownian time units. 
During this time-window the mesoparticle could be regarded as a patchy particle, which would 
surely exhibit anisotropic interactions with neighboring meso-particles. 

\begin{figure}
\begin{center}
\includegraphics[width=0.5\textwidth]{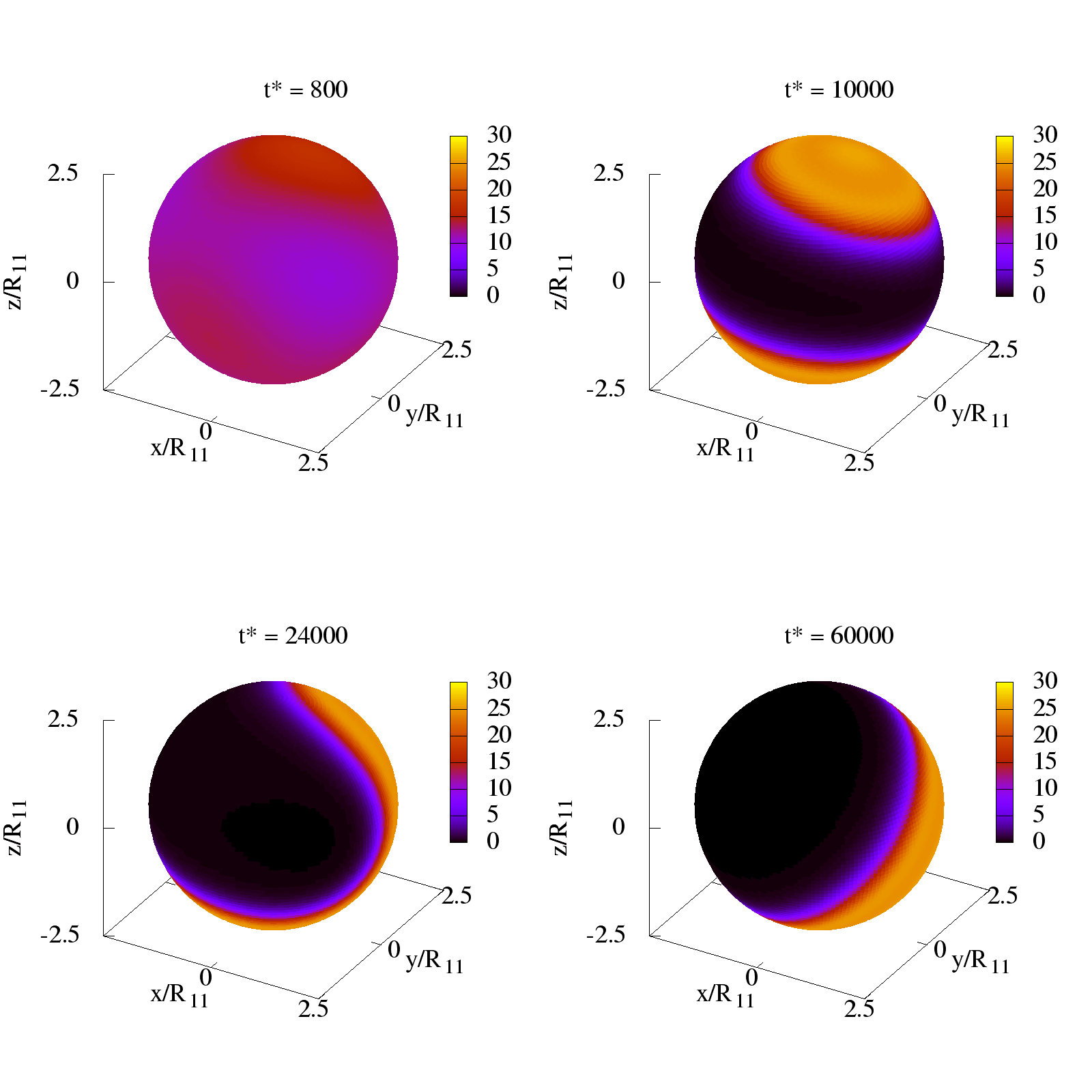}
\caption{
{\bf Smaller sphere with $\mathbf{x=0.5}$}.
Time-evolution of the density with 
$\rho R_{11}^2 = 25$ on a meso-particle with radius $R=2.5 R_{11}$.
The surface particles become trapped in a `banded' metastable state.  
For comparison, the density on a larger meso-particle 
(see Fig.\ref{fig:spinodal_decomposition_x_0.5}) at the same state-point does not 
display such a banded structure at any point during the time-evolution. 
}
\label{fig:spinodal_decomposition_x_0.5_R2.5}
\end{center}
\end{figure}



\begin{figure}
\begin{center}
\includegraphics[width=0.5\textwidth]{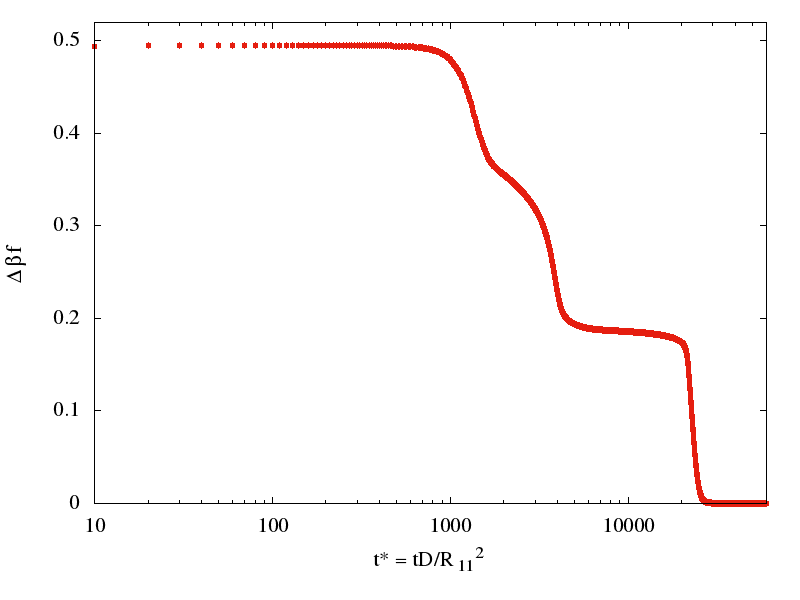}
\caption{
{\bf Smaller sphere with $\mathbf{x=0.5}$}.
Time-evolution of the free energy for statepoint $\rho R_{11}^2 = 25$ on a meso-particle with 
radius $R=2.5 R_{11}$.
The sudden decrease in the free energy at around $t^*=20000$ corresponds to the breaking of density 
`band' around the meso-particle (see figure \ref{fig:spinodal_decomposition_x_0.5_R2.5}). 
}
\label{fig:free_energy_x_0.5_R2.5}
\end{center}
\end{figure}

We next consider phase separation on a smaller meso-particle, $R=2.5R_{11}$, for which finite-size 
effects become important. 
A typical example of the time-evolution of the density is shown in 
Fig.~\ref{fig:spinodal_decomposition_x_0.5_R2.5} and the corresponding free energy in 
Fig.~\ref{fig:free_energy_x_0.5_R2.5}, where we address first the statepoint A in the phase diagram 
($\rho\!=\!25$, $x\!=\!0.5$). 
When compared with the phase separation dynamics on the larger meso-particle 
(see Fig.~\ref{fig:spinodal_decomposition_x_0.5}) we observe from the decay of the free energy that, 
although the onset time of the initial instability is larger for the smaller sphere, the overall time 
taken to arrive at the equilibrium state is smaller. 

In contrast to the behavior on the larger meso-sphere, the density evolves here into a `band' state, 
where two islands with species 1 form, separated by a band of species 2 particles. 
This state is stable over a long time, which can be seen in the plateau of the free energy 
(from $t^* \sim 5000$ to $t^* \sim 20000$), 
before it finally collapses to reach an equilibrium state qualitatively similar to that found 
on the larger sphere. 
The interesting feature here is that, despite the symmetric composition ($x\!=\!0.5$), the 
time-evolution is qualitatively closer to Ostwald ripening than classic spinodal decomposition.  
This is a finite-size effect, which arises because `long wavelength modes' (a notion to be clarified 
in the following section) are suppressed by the relatively small circumference of the meso-sphere, relative 
to the size of the surface particles.


%

\begin{figure}
\begin{center}
\includegraphics[width=0.5\textwidth]{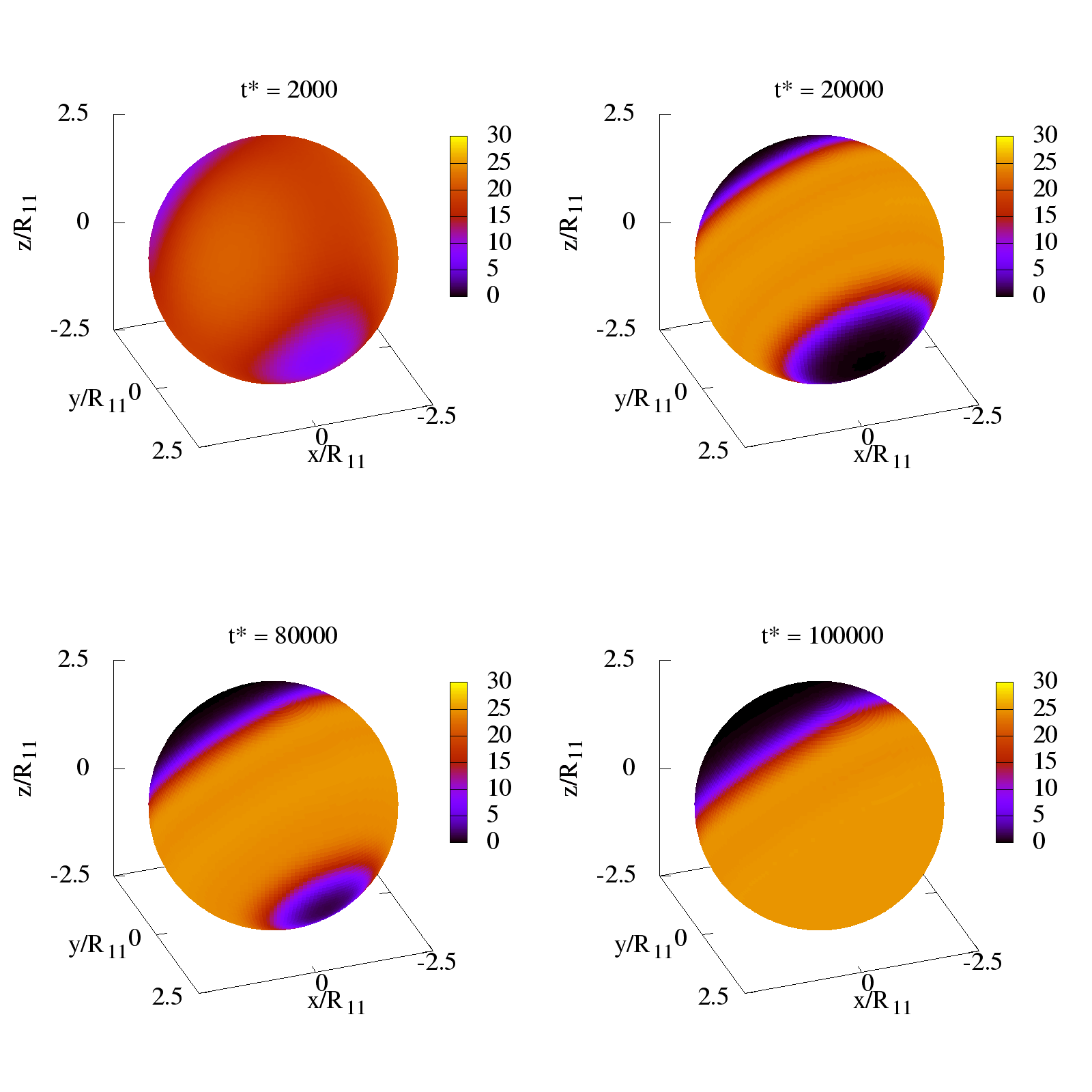}
\caption{
{\bf Smaller sphere with $\mathbf{x\!=\!0.3}$}. 
Time-evolution of the density with $\rho R_{11}^2 = 25$ on a meso-sphere of radius $R=2.5 R_{11}$.
The banded state is here more stable than in the case $x\!=\!0.5$ 
(see Fig.~\ref{fig:spinodal_decomposition_x_0.5_R2.5}).
}
\label{fig:spinodal_decomposition_x_0.3_R2.5}
\end{center}
\end{figure}

\begin{figure}
\begin{center}
\includegraphics[width=0.5\textwidth]{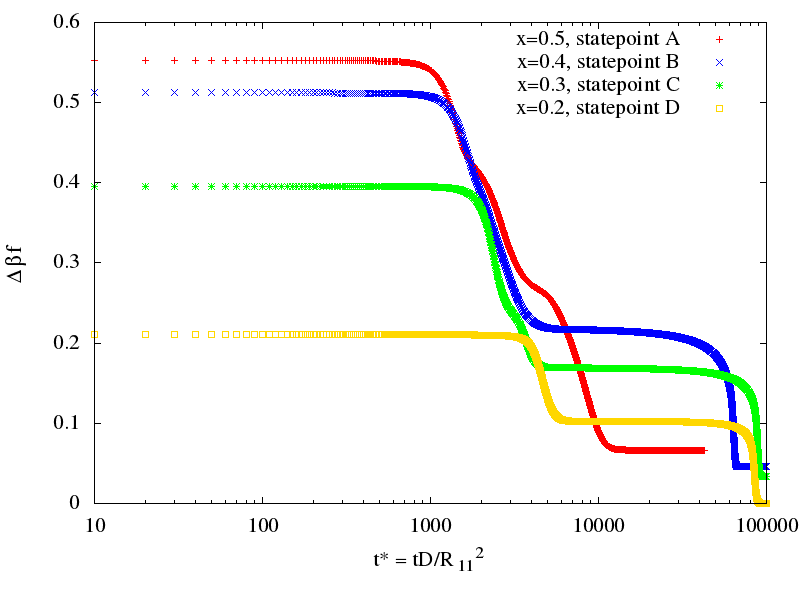}
\caption{
{\bf Smaller sphere with $\mathbf{x\!=\!0.2\ldots0.5}$}.
Time-evolution of the free energy for $\rho R_{11}^2 = 25$ on a meso-sphere of radius $R=2.5R_{11}$. 
The plateau in the free energy corresponds to the banded state.  
The lifetime of the density band increases as $x$ is reduced. 
}
\label{fig:free_energy_R2_5_comparison_in_x}
\end{center}
\end{figure}

If the value of $x$ is reduced for fixed $\rho$, the statepoint moves towards the spinodal and the time 
taken for the system to reach equilibrium increases. 
In Fig.~\ref{fig:spinodal_decomposition_x_0.3_R2.5} we show an example of the density evolution for 
the value $x\!=\!0.3$ (statepoint C). 
We again observe the formation of a band around the particle, however, this metastable state is much 
longer-lived than that observed for the case $x\!=\!0.5$, as can be seen from the time-evolution of the 
free energy shown in Fig.~\ref{fig:free_energy_R2_5_comparison_in_x}. 
In general, we find that the smaller the value of $x$, the more stable the band structure becomes. 
In Fig.~\ref{fig:free_energy_R2_5_comparison_in_x} we show the 
free energy per particle as a function of time 
for statepoints A, B, C and D in figure \ref{fig:binodal_volume}, corresponding to 
$x=0.5, 0.4, 0.3$ and  $0.2$. 
This enhanced stability of the band structure can be attributed to the fact that the distance 
between the interfaces increases as the surface coverage of the minority phase is reduced by reducing 
$x$.



%



The process of spinodal decomposition in bulk systems is commonly 
subdivided into different dynamical 
regimes.
In the early stages of phase separation density gradients are small and the dynamics can be 
well described using Cahn-Hillard theory 
(see e.g., Refs.~\cite{Dhont1996, Evans1979a, Abraham1976, Archer2004a}).
Early-stage spinodal decomposition is characterized by an exponential growth of low-wavelength 
density fluctuations \cite{Archer2004a}. 
For infinite, flat systems the fluctuation spectrum is conveniently analyzed 
using the Fourier transform, which enables unstable wavenumbers $k$ to be identified. 
In the present situation, where the surface fluid is confined to a spherical surface of finite extent, 
the analogue of the wavenumber is provided by the $l,m$ labels of the spherical harmonic expansion 
of the density field.   

We define the early-stage of spinodal decomposition to be the time-window following the quench, for 
which the linearized theory agrees with a full non-linear calculation. 
Deviations indicate the onset of intermediate-stage phase separation. 
We thus follow \cite{Archer2004a} and linearize the DDFT equation in the density fluctuation 
$\tilde{\rho}_i(\mathbf{r},t) = \rho_i(\mathbf{r},t) - \rho_i^b$. We first express the DDFT equation 
in the form
\begin{eqnarray}
\label{eq:ddfteq_tilde}
 \beta \Gamma^{-1} \frac{\partial \tilde{\rho}_i(\mathbf{r},t)}{\partial t} &=& \nabla^2 \tilde{\rho}_i(\mathbf{r},t) - \nabla \bigl(\tilde{\rho}_i(\mathbf{r},t) \nabla c^{(1)}_i(\mathbf{r},t) \bigr) \nonumber \\
 &-& \rho_i^b \nabla^2 c^{(1)}_i(\mathbf{r},t).
\end{eqnarray}
and substitute into this expression a functional Taylor expansion of the one-body direct correlation function
\begin{eqnarray}
 \hspace*{-0.4cm}c_i^{(1)}(\mathbf{r},t) &=& 
 \sum_{j} c_i^{(1)}(\mathbf{r},t) \Big|_{\rho_j^b} \notag\\
 &+& \sum_j \int d\mathbf{r}' c_{i,j}^{(2)}(\mathbf{r},\mathbf{r}',t) \Big|_{\rho_j^b} \, \tilde{\rho}_j(\mathbf{r}',t) + ...
\end{eqnarray}
For the GCM surface fluid this yields to first order in density fluctuations the following result
\begin{eqnarray}\label{expansion}
\beta^{-1} c_1^{(1)}(\mathbf{r},t) &=& -\rho_1^b \hat{v}_{11} - \rho_2^b \hat{v}_{12} - \tilde{\rho}_1 * v_{11} - \tilde{\rho}_2 * v_{12}, 
\notag\\
\beta^{-1} c_2^{(1)}(\mathbf{r},t) &=& -\rho_2^b \hat{v}_{22} - \rho_1^b \hat{v}_{12} 
- \tilde{\rho}_2 * v_{22} - \tilde{\rho}_1 * v_{12},\notag\\
\end{eqnarray}
where the star denotes a convolution. 
Substitution of \eqref{expansion} into \eqref{eq:ddfteq_tilde} and retaining linear terms yields 
\begin{align}
 \beta \Gamma^{-1} \frac{\partial \tilde{\rho}_1(\mathbf{r},t)}{\partial t} &=& \nabla^2 \tilde{\rho}_1(\mathbf{r},t) - \beta \rho_1^{b} \nabla^2(v_{11} * \rho_1 + v_{12} * \rho_2), 
\label{linear1}\\ 
 \beta \Gamma^{-1} \frac{\partial \tilde{\rho}_2(\mathbf{r},t)}{\partial t} &=& \nabla^2 \tilde{\rho}_2(\mathbf{r},t) - \beta \rho_2^{b} \nabla^2(v_{22} * \rho_2 + v_{12} * \rho_1).
\label{linear2}
\end{align}

To identify the regime of early-stage phase separation we have compared 
the free energy from the nonlinear DDFT, Eq.~\eqref{stDDFT}, with the results obtained by solving 
Eqs.~\eqref{linear1} and \eqref{linear2}.  
%
%
%
Our numerical calculations show where the linearized solution begins to 
deviate from the full solution of the DDFT equations.   
%
%
For a given meso-particle radius we evaluate the density field at a time just prior to this 
deviation, determine the coefficients in the spherical harmonic expansion, Eq.~\eqref{harmonic_expansion}, 
and then average over the $m$-index. 
Furthermore, we average over a set of $50$ different initial conditions. 
The resulting averaged coefficient, $\langle a_l^m \rangle$, is a function of the index $l$  
and indicates which modes of the density field contribute most to  the density instability.    
 
\begin{figure}
\begin{center}
\includegraphics[width=0.5\textwidth]{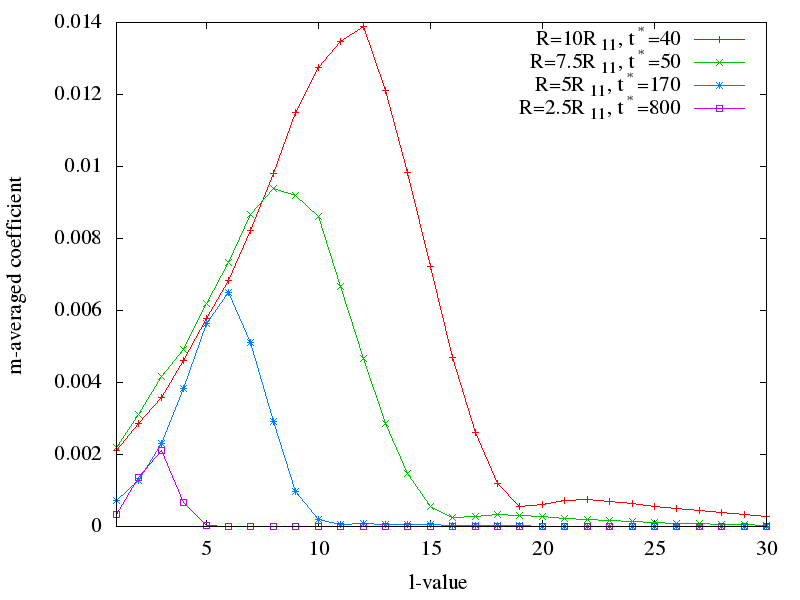}
\caption{$l$-modes of the early-stage spinodal decomposition for different meso-particle sizes.
We show the spherical harmonic coefficient for a given $l$-value, averaged over the $m$ indices 
from $m\!=\!-l\ldots l$ and averaged over initial conditions.
For smaller sphere sizes the peak shift to smaller $l$-values.
}
\label{fig:early_stage_coefficients}
\end{center}
\end{figure}

In Fig.~\ref{fig:early_stage_coefficients} we show $\langle a_l^m \rangle$ as a function 
of $l$ for different meso-particle radii. 
For the familiar case of spinodal decomposition in a flat space, it is standard procedure 
to analyze the static structure factor in order to identify unstable Fourier modes. 
In the present situation, where a liquid is constrained to lie on a finite spherical surface, 
the $\langle a_l^m \rangle$ data shown in Fig.~\ref{fig:early_stage_coefficients} provide 
an appropriate analogue to the structure factor.  
%
%
%
%
%
For smaller sphere sizes the dominant $l$-values are lower than for larger spheres.
An explanation for this effect is that the wavelengths which dominate the instability, 
the `ripples' on the sphere surface, are independent of the sphere size and, therefore, 
on smaller spheres are described by a smaller $l$-value.
Using Jeans' rule one can identify the wavelength of a spherical harmonic with 
degree $l$ by $\lambda = 2 \pi R /(l+1/2)$.

\begin{figure}
\begin{center}
\includegraphics[width=0.5\textwidth]{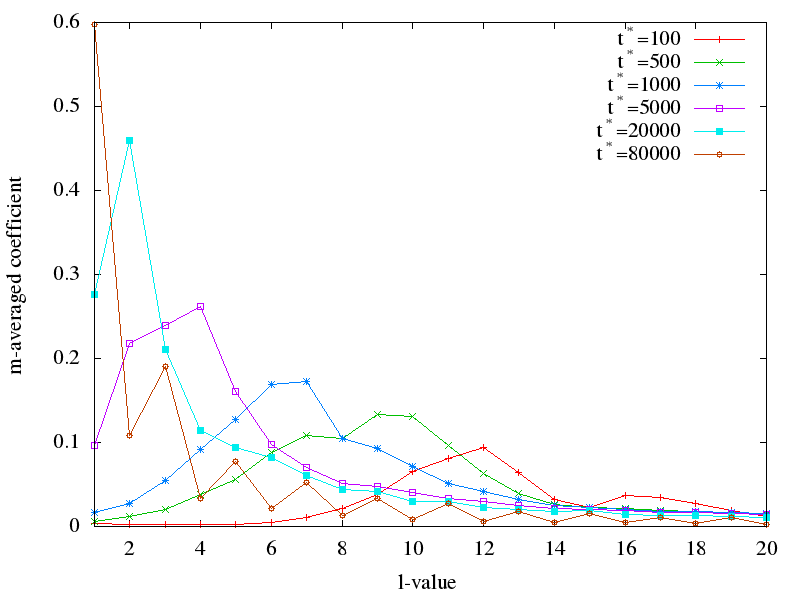}
\caption{l-Modes for different times for a meso-particle of radius $R=10R_{11}$. 
As the density evolves the smaller $l$-modes give an increasing contribution.   
In the final state the dipole dominates.}
\label{fig:evolution_coefficients}
\end{center}
\end{figure}

As the density relaxes from its initial to its final state the $\langle a_l^m \rangle$ 
evolve in time.
For a sphere of radius $R\!=\!10R_{11}$ and $x\!=\!0.5$ we show in 
Fig.~\ref{fig:evolution_coefficients} this time-evolution from the end of early-stage 
spinodal decomposition, all the way to the final state. 
For times just beyond the early-stage of spinodal decomposition we observe the same behaviour as 
seen in Fig.~\ref{fig:early_stage_coefficients}. 
However, the peak of the curve shifts to smaller values as time increases. 
In the final state the surface fluid is completely phase-separated and the dominant mode is the 
dipole ($l\!=\!1$).
In all plots we excluded the $l=0$ contribution, which only represents a 
homogeneous field and has therefor no contribution to the angular distribution.
%



\begin{figure}
\begin{center}
\includegraphics[width=0.5\textwidth]{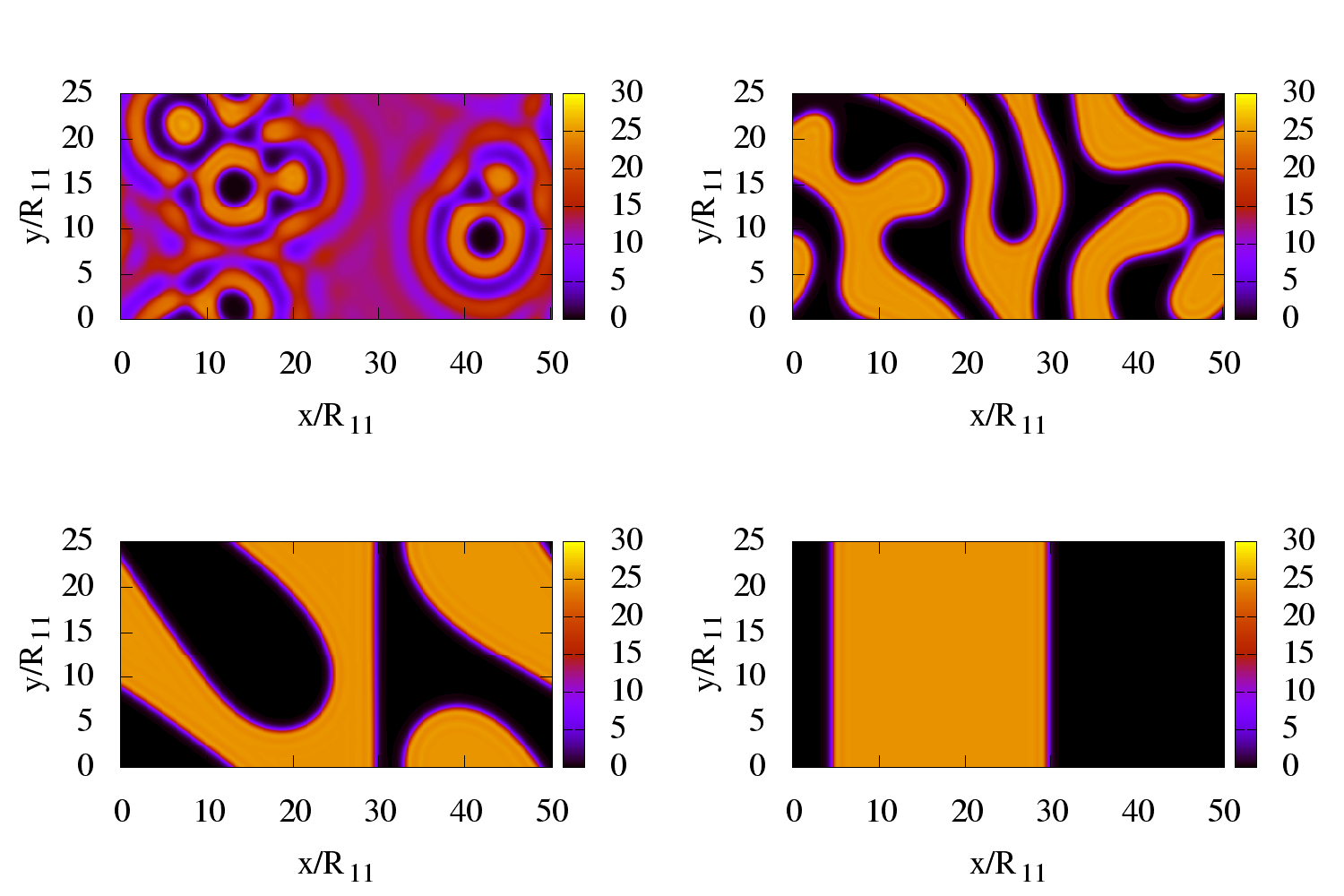}
\caption{Time evolution of the density distribution for a Gaussian mixture confined to a flat surface. 
The system has periodic boundary conditions, to insure the conservation of particles. The number of particles/surface area is equivalent to the sphere system with radius $R=10 R_{11}$.
We see that the dynamics on the flat grid appear to be much faster, since the equilibrium state is reached after roughly $t^*=7000$, which took $t^*=60000$ on the sphere (see figure \ref{fig:spinodal_decomposition_x_0.5}).
The corresponding evolution of the free energy can be seen in figure \ref{fig:free_energy_x_0.5_flat}
}
\label{fig:spinodal_decomposition_x_0.5_flat}
\end{center}
\end{figure}

\begin{figure}
\begin{center}
\includegraphics[width=0.5\textwidth]{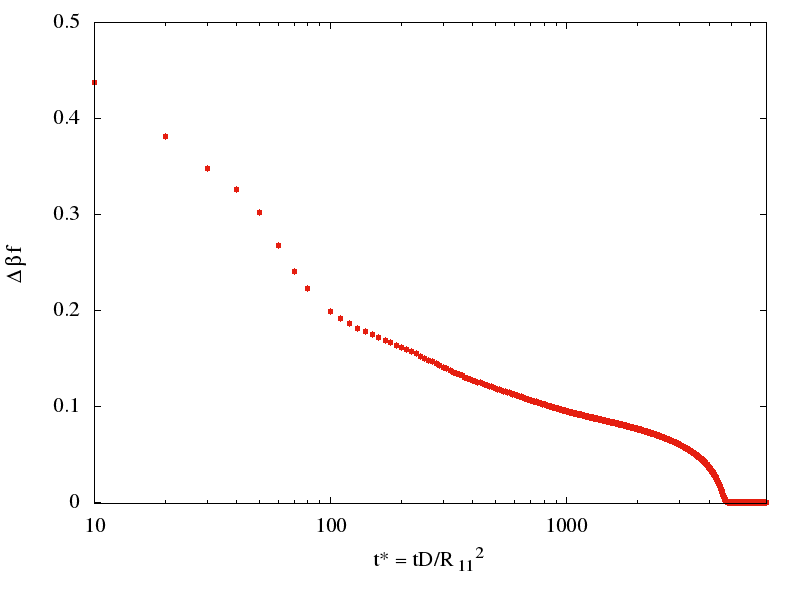}
\caption{Free energy time evolution for the spinodal decomposition on a flat surface with periodic boundary conditions. The flat system shows qualitatively similar behavior to the spherical system.
The free energy shows a significant drop in the initial stages of spinodal decomposition and afterwards slowly decreases as the system reaches its equilibrium state.}
\label{fig:free_energy_x_0.5_flat}
\end{center}
\end{figure}

The dynamics of phase separation on a 
meso-particle can be compared with phase separation in a 
flat, planar system.
For this comparison we use the same parameters $R_{ij}$ and $\epsilon_{ij}$ as previously and employ 
periodic boundary conditions. 
The number of particles is set equal in the flat and curved systems. 
%
%
The time evolution of the density and free energy for the flat system are shown in 
Figs.~\ref{fig:spinodal_decomposition_x_0.5_flat} and \ref{fig:free_energy_x_0.5_flat}, respectively. 
In both cases we use $x\!=\!0.5$. 
The main observation is that the dynamics of spinodal decomposition are much faster for the flat system than 
the corresponding spherical system. 
The equilibrium state is reached after approximately $t^*=7\times 10^3$, compared to the curved system which 
took $t^*=6\times 10^4$, an order of magnitude longer, to achieve comparable equilibration 
(see figure \ref{fig:spinodal_decomposition_x_0.5}).

From Fig.~\ref{fig:spinodal_decomposition_x_0.5_flat} it is apparent that the periodic boundary conditions 
artificially constrain the orientation of the interface.
This unphysical constraint is absent on the sphere, since its topology does not need any boundary 
conditions and the interfaces can have arbitrary orientation. 
From our calculations it would appear that the finite-size effects associated with smaller meso-spheres 
have a stabilizing effects on the band structure.



%

\section{Interacting meso-particles}

Going beyond the dynamics of phase separation on a single sphere, we next investigate 
the interaction between a pair of meso-spheres. If two meso-spheres are sufficiently 
close that their surface particles interact, then they exhibit an anisotropic 
interaction. Understanding the pair interaction can then form a basis for 
investigating the structures which may result from self-assembly.   
%


Calculating the interaction potential between two meso-particles requires as input the 
distance between two arbitrary points, one located on meso-particle 1 and the other on 
meso-particle 2. 
For convenience we fix meso-particle 1 at the origin of a cartesian coordinate system 
(henceforth referred to as the `left particle', with radius $R_L$).  
The center of the right particle (radius $R_R$) is chosen to lie on 
the positive $x$-axis. 
The center-to-center distance is $R_L + R_R + d$, and if $d$ is comparable to the range of the 
Gaussian interaction $R_{ij}$, then the two particles will influence each other.

The distance $|z|$ between any point $(\phi_L, \theta_L)$ on the surface of the left sphere and any 
point $(\phi_R, \theta_R)$ on the surface of the right sphere is given by
\begin{eqnarray*}
|z|^2\! &=&\! (R_L \sin\theta_L \cos\phi_L - R_R \sin\theta_R \cos\phi_R - d - R_L - R_R)^2 \\
&+& (R_L \sin\theta_L \sin\phi_L - R_R \sin\theta_R \sin\phi_R)^2 \\
&+& (R_L\cos \theta_L - R_R\cos\theta_R)^2.
\end{eqnarray*}
The external potential exerted on particle species $i=1,2$ on the left sphere by the right sphere is 
thus given by
\begin{eqnarray}
\label{twin_external}
 \beta V_{ext}(\theta,\phi)_{Li} &=& \beta \int d\Omega' \rho_{R1}(\Omega')\, 
v_{i1}\bigl(|z|(\Omega,\Omega')\bigr)
\\
&+& \beta \int d\Omega'  \rho_{R2}(\Omega')\, v_{i2}\bigl(|z|(\Omega,\Omega')\bigr).
\notag
\end{eqnarray}
The external potential acting on the right sphere is then simply obtained by exchanging the labels 
$R$ and $L$ in the above expression.
 %
%
%

In the numerical time integration of the DDFT equation it is expensive to compute these 
integrals at each time step. 
In principle, to solve the time-evolution in a fully self-consistent way, the density on each particle 
surface should be subject at each time-step to the instantaneous external field generated by the 
density distribution on the surface of the other sphere. 
However, a fully self-consistent solution seems to us to be unnecessary. 
The two meso-particles are mobile objects and, provided the density is not too high, the process 
of phase separation on each meso-particle will largely proceed in the absence of significant interaction 
with the others. 
From our single meso-particle studies we have shown that the patchy domain structure can be a long-lived 
metastable state. 
It is thus rather likely that meso-particles which drift together and interact do so while trapped 
in a metastable state. 
More precisely, we assume that the timescale of collisions between meso-spheres, 
$D_m^{-1}\rho_m^{-2/3}$, where $D_m$ and $\rho_m$ are the diffusion constant and density of the 
meso-spheres, is less than the lifetime of the metastable states on the individual meso-sphere surfaces.

Due to the above considerations we can simplify the problem by considering the interaction of 
meso-spheres with static surface density distributions. 
These static distributions are obtained from the single particle calculations presented in 
section \ref{results}.
For a given interparticle separation we seek the lowest energy relative orientation of a pair of 
meso-spheres. 
Using Eq.~\ref{twin_external} we can determine for all relative orientations 
the potential acting upon each meso-sphere due to its neighbor and, thus, the dependence of 
the total free energy on the relative orientation and separation of the meso-sphere pair. 
In Appendix 3 we report the techniques required for this calculation.
 
\begin{figure}
\begin{center}
\includegraphics[width=0.5\textwidth]{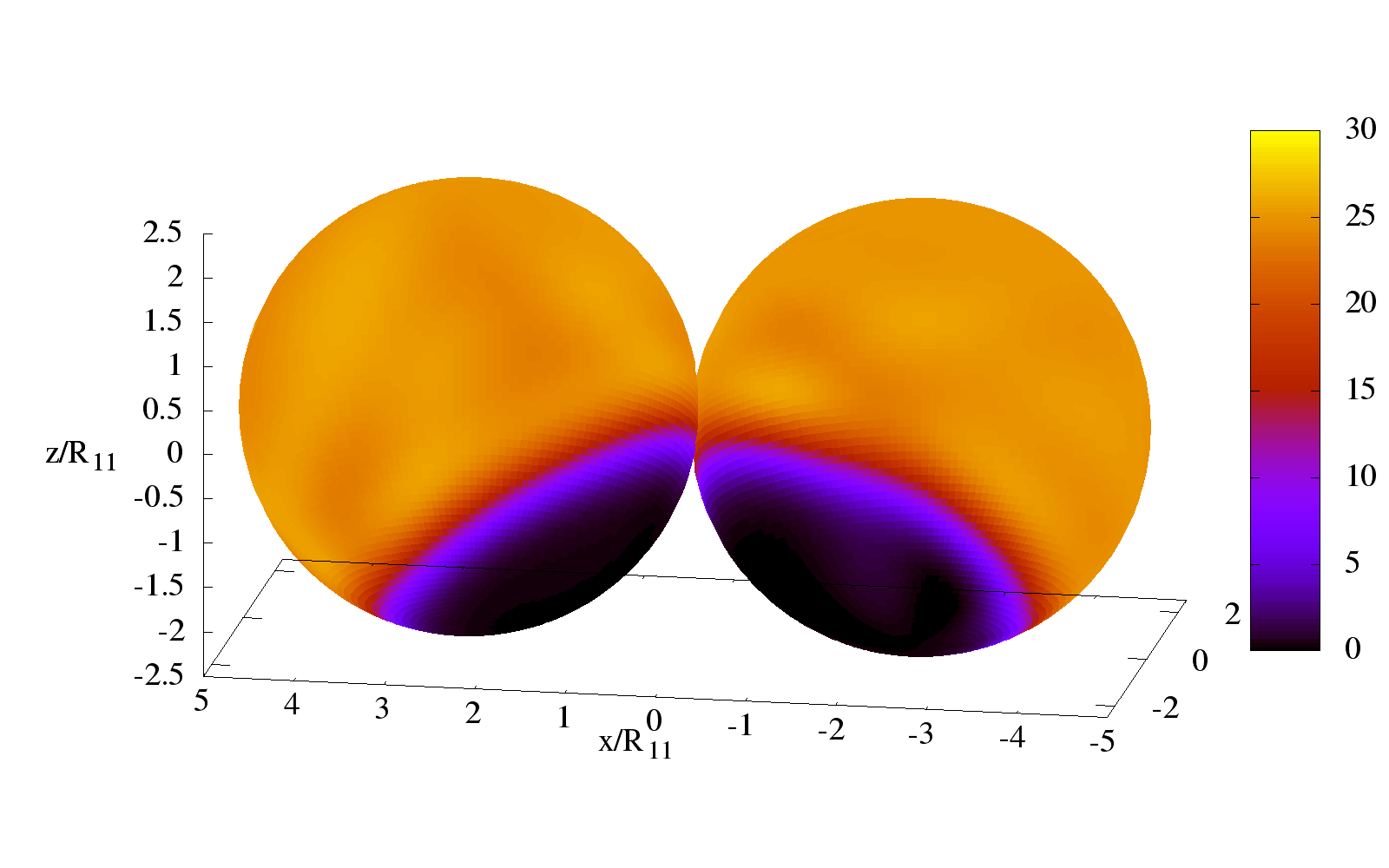}
\caption{Two interacting meso-spheres with radius $R=2.5R_{11}$ and $x=0.2$. 
The configuration shown minimizes the total free energy.
}
\label{fig:twins_x_0_2}
\end{center}
\end{figure}

\begin{figure}
\begin{center}
\includegraphics[width=0.5\textwidth]{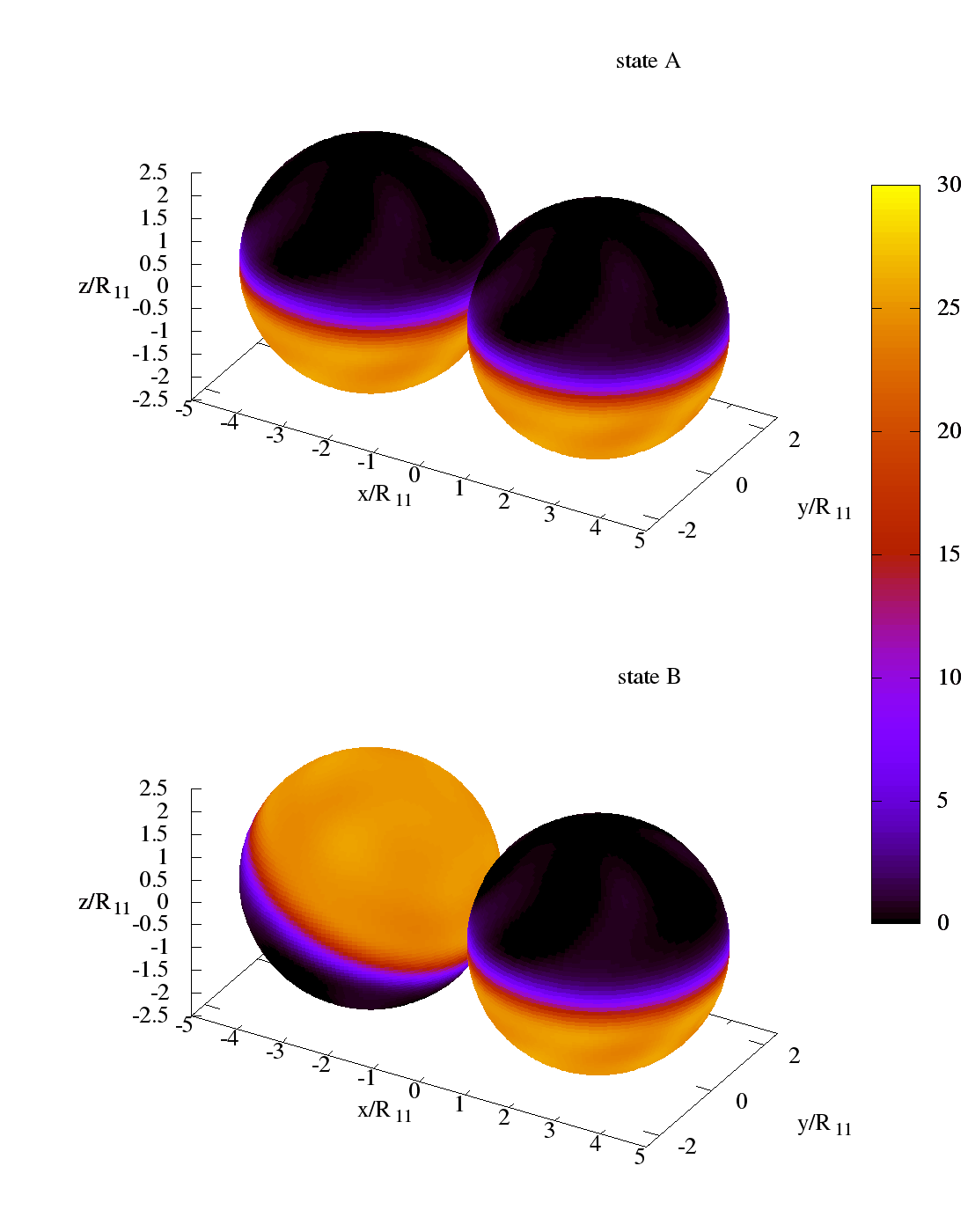}
\caption{Two interacting spheres with mixing parameter $x=0.5$. 
The plot shows two different configurations A and B. State A is the configuration with a minimal energy cost. Turning one of the spheres away from this configuration (state B) leads to an extra energy cost.
The energy cost as a function of angle $\theta$ is shown in figure \ref{fig:two_sphere_interaction}}
\label{fig:twins_x_0_5}
\end{center}
\end{figure}

\begin{figure}
\begin{center}
\includegraphics[width=0.45\textwidth]{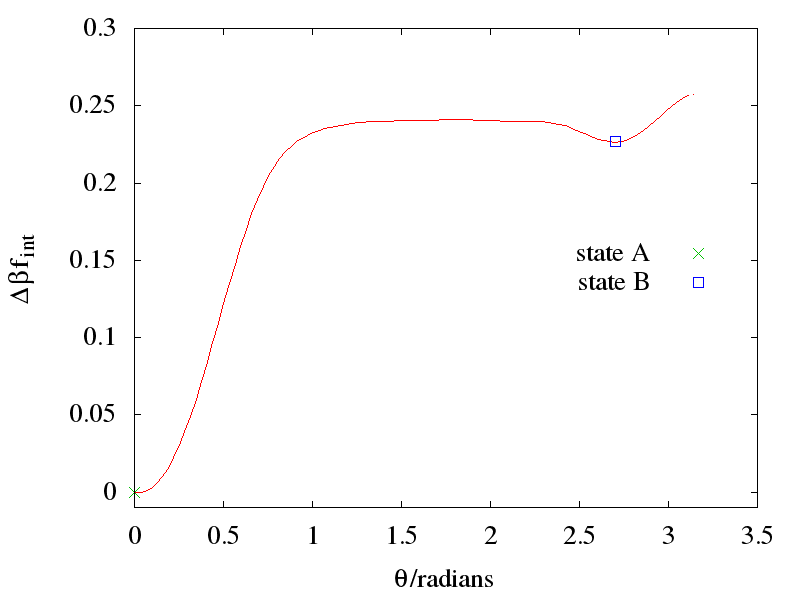}
\caption{The plot shows the energy cost per particle as a function of angle, when turning one of the two interacting spheres away from its equilibrium configuration.}
\label{fig:two_sphere_interaction}
\end{center}
\end{figure}

For simplicity we will limit ourselves to the interaction between meso-spheres for 
which the phase separation process is fully completed. 
In Fig.~\ref{fig:twins_x_0_2} we show the configuration of minimum free energy 
for $R=2.5R_{11}$ and $x=0.2$. 
In this case the spheres orient such that the interfaces between domains are touching. 
The choice of mixing parameter $x$ thus specifies the 
`bond angle' between the two meso-spheres. 
In Fig.~\ref{fig:twins_x_0_5} we show two different configurations of meso-spheres with 
radius $R=2.5R_{11}$, but now for $x=0.5$.
The first configuration shown (state A) yields the lowest value of the free energy. 
By flipping one of the spheres (state B) we obtain a state with higher free energy, but 
which represents a local minimum in the free energy. 
In Fig.~\ref{fig:two_sphere_interaction} we show the dependence of the free energy on the 
angle $\theta$ (Euler angle for rotation around the $y$-axis, see also Appendix 3).

We would like to emphasize that, for the present GCM surface particles, the 
interaction forces acting between meso-spheres are repulsive. The minimum free energy 
configurations identified here correspond to situations of minimal repulsion for a given 
particle separation. While this is somewhat different from the standard picture of 
synthetic patchy particles (for which the patches are mutually attractive) we expect 
the anisotropic repulsion presented by the present model to be important for determining 
the packing structure of the meso-spheres at intermediate and high densities.

The configurations shown in Figs.~\ref{fig:twins_x_0_2} and \ref{fig:twins_x_0_5}, together with 
the free energy in Fig.~\ref{fig:two_sphere_interaction}, indicate that fully phase separated 
meso-spheres will show interesting self-organization behavior, which can be tuned by 
varying the value of the mixing parameter $x$. 
For $x=0.5$ it is clear from the minimum energy state A (shown in Fig.~\ref{fig:twins_x_0_5}) that 
an assembly of many phase separated meso-spheres would build sheets of particles with hexagonal 
in-plane packing. 
Indeed, precisely this behaviour was found in computer simulations of a closely related 
model of patchy particles \cite{Zhang2004}. In this study the authors considered the self assembly 
of hard spheres with discrete attractive patches positioned around the equator. We thus anticipate 
that our particles with $x\!=\!0.5$ will show very similar self-assembly. A distinction between our 
model and that studied in \cite{Zhang2004} is that our meso-particles do not possess an 
up-down symmetry. 
Our minimum energy state would have all meso-particles oriented in the 
same direction, however, the fact that the `flipped state' (state B in Fig.~\ref{fig:twins_x_0_5}) is a local free 
energy minimum, suggests that a certain fraction of the meso-particles in the sheet will be 
flipped with respect to the majority.  

For $x\ne 0.5$ the bond angle is no longer zero. In a system of many particles this would lead 
in general to a `buckled' sheet of particles which would be subject to geometrical frustration 
effects. However, for particular choices of $x$ the bond angle can be made compatible with a closed 
shell of particles. 
The findings of Ref.~\cite{Zhang2004} support this speculation; simulations were performed on systems 
of hard-spheres with a ring of discrete attractive patches lying away from the equator.  

%


\section{Conclusions}\label{conclusions}

In this paper we have studied the process of phase separation on the surface of 
a sphere using the method of dynamical density functional theory with a simple 
mean-field free energy functional. 
For larger meso-sphere radii we find standard spinodal decomposition dynamics 
for an equal mixture, $x=0.5$, leading to a `half-half' final state. As the value 
of $x$ is reduced towards the spinodel, then the phase separation dynamics are 
given by the Ostwald ripening scenario, as expected. The 
long-lived metastable states, consisting of islands of minority phase, could 
behave as patchy particles with potentially interesting self-assembly properties. 
An unexpected finding is that smaller meso-particles do not exhibit typical spinodal 
decomposition dynamics for any value of $x$. Even for the symmetric mixture with $x=0.5$ 
the phase separation resembles Ostwald ripening. 

For the case of a fully phase-separated larger sphere we have considered the interaction 
between pairs of meso-particles in order to gain insight into possible self-assembly 
mechanisms. 
For a pair of meso-particles in contact with each other we find the state of minimum free 
energy to be that where the interfaces between domains are touching 
(see Figs.~\ref{fig:twins_x_0_2} and \ref{fig:twins_x_0_5}) and the meso-particles have 
the same orientation. The state for which the particles have opposite orientation is 
a less favorable metastable minimum of the free energy. 

In Ref.~\cite{Zhang2004} the self-assembly of a simplified version of our phase-seperated meso-particles
has been studied. Simulations were performed of particles with discrete attractive interaction 
sites at fixed locations on the particle surface.
For particles with an attractive ring-like patch around the equator, self-assembly into particle 
sheets was identified. 
When the ring of discrete sites was displaced from the equator then the sheets became bent and 
frustrated. 
From our findings, it would appear that a fully phase-separated binary mixture on the surface of 
each meso-particle provides an approximate realization of the toy model simulated in 
Ref.~\cite{Zhang2004}. The self-assembly properties can thus be controlled by varying the 
mixing parameter $x$ of the surface particles. 
%
%
One can thus speculate about the more complex structures which could arise when meso-particles 
in metastable states (e.g.~that shown in Fig.~\ref{fig:spinodal_decomposition_x_0.3}) interact 
with each other. 
We plan to perform extensive Brownian dynamics computer 
simulations of simplified models to investigate the self-organized structures which can develop 
in these systems. 

Finally, we note that there have been experimental observations on the formation of stripe 
patterns formed by immiscible ligands coadsorbed on the surface of gold and silver nanoparticles 
\cite{Jackson2004}.
Supporting atomic simulation studies have shown similar stripe formation for surfactants on spherical 
surfaces \cite{Singh2007}. It would be interesting to see if such structured are captured by 
the simple density functional approach employed in the present study.

\section*{Acknowledgements}
This research was supported by the Swiss National Science Foundation
through the National Centre of Competence in Research Bio-Inspired Materials.


\appendix

\section*{Appendix 1}\label{appendix1}
We here recall the conditions for phase coexistence in the binary mixture \cite{Louis2000}.
The thermodynamic stability conditions are given by
\begin{subequations}
\label{eq:condition}
 \begin{align*}
  \biggl(\frac{\partial^2 f}{\partial v^2} \biggr)_x &> 0 \\
  \biggl(\frac{\partial^2 f}{\partial x^2} \biggr)_v &> 0 \\
  \biggl(\frac{\partial^2 f}{\partial v^2} \biggr)_x   \biggl(\frac{\partial^2 f}{\partial x^2} \biggr)_v -  \biggl(\frac{\partial^2 f}{\partial v \partial x} \biggr)^2 &> 0, 
 \end{align*}
\end{subequations}
where $f$ is the Helmholtz free energy per particle and $v=\rho^{-1}$.
The first inequality ensures mechanical stability (positive compressibility), the second inequality 
is the condition against spontaneous demixing at constant volume
and the final inequality ensures stability at constant pressure. 
With the free energy density from equations 
\eqref{ideal_perparticle} and \eqref{excess_perparticle} the stability conditions 
can be reduced to
\begin{subequations}
\label{eq:condition2}
 \begin{align}
   1 + \rho \hat{V}_0(x) &> 0 \label{eq:con1} \\
   1- \rho x (1-x)\chi &> 0 \label{eq:con2} \\
   1 + \rho \hat{V}_1(x) - \rho^2 x (1-x) \Delta &> 0, \label{eq:con3}
 \end{align}
\end{subequations}
where we have defined the following parameters
\begin{eqnarray*}
  \chi &=& 2 \hat{v}_{12} - (\hat{v}_{11} + \hat{v}_{22}) \\
  \Delta &=& \hat{v}_{12}^2 - \hat{v}_{11}\hat{v}_{22} \\
  \hat{V}_1(x) &=&  (1-x)\hat{v}_{11} + x \hat{v}_{22}
\end{eqnarray*}

The first inequality (\ref{eq:con1}) is always fulfilled for the Gaussian interaction, since $\hat{V}_0(x)$ is strictly positive.
Phase separation is possible provided that condition (\ref{eq:con2}) or condition (\ref{eq:con3}) are violated.
Below we consider phase coexistence at constant volume, resulting in violation of condition (\ref{eq:con2}).

\subsection{Phase Separation at Constant Volume}
\label{sec:constant_volume}
Violation of condition (\ref{eq:con2}) requires $\chi > 0$
\begin{equation}
\label{eq:chi1}
 \chi = \pi \bigl[2 \epsilon_{12}^* R_{12}^2 - (\epsilon_{11}^* R_{11}^2 + \epsilon_{22}^* R_{22}^2)\bigr] > 0
\end{equation}
Whether phase separation is possible or not depends on the choice of the parameters $\epsilon_{ij}^*$ and $R_{ij}$.
From equation (\ref{eq:chi1}) we see a simple choice is $R_{11} = R_{22} = R_{12}$ and $\epsilon_{12}^* > \epsilon_{11}^* = \epsilon_{22}^*$.
For this choice of paramters it is physically intuitive that the system might phase separate, as the energy penalty for unlike particles being close to each other is higher then for alike particles.

The physically instable region of the phase diagram is given by stability condition (\ref{eq:con2}). Instability occurs first, when 
\begin{equation*}
1 -\rho x (1-x)\chi = 0.
\end{equation*}
Thus the spinodal line is given by
\begin{equation}
 \rho_s(x) = \frac{1}{x(1-x)\chi}.
\end{equation}
The binodal (phase coexistence line) is determined by chemical and mechanical equilibrium.
This means that the chemical potential of both particle species (1 and 2), as well as the pressure is equivalent in both phases (A and B):
\begin{eqnarray}
\label{eq:coexcondition}
\mu_1(\rho,x_A) &=& \mu_1(\rho,x_B), \nonumber \\
\mu_2(\rho,x_A) &=& \mu_2(\rho,x_B), \\
p(\rho,x_A) &=& p(\rho,x_B). \nonumber
\end{eqnarray}
Chemical potential and pressure are obtained from the free energy density via:
\begin{eqnarray*}
 \mu_1 &=& f - v \biggl(\frac{\partial f}{\partial v}\biggr)_x - x \biggl(\frac{\partial f}{\partial x}\biggr)_v, \\
 \mu_2 &=& f - v \biggl(\frac{\partial f}{\partial v}\biggr)_x + (1-x) \biggl(\frac{\partial f}{\partial x}\biggr)_v,\\
 p &=& -\biggl(\frac{\partial f}{\partial v}\biggr)_x.
\end{eqnarray*}
After simplification one finds:
\begin{eqnarray}
 \beta \mu_1 &=& \ln \bigl(\rho \lambda^2 (1-x) \bigr) + \rho (1-x)\hat{v}_{11}(0) + \rho x \hat{v}_{12}(0), \\
 \beta \mu_2 &=& \ln \bigl(\rho \lambda^2 x \bigr) + \rho (1-x)\hat{v}_{12}(0) + \rho x \hat{v}_{22}(0), \\
 \beta p &=& \rho + \frac{1}{2} \rho^2 \hat{V}_0(x), \label{eq:pressure}
\end{eqnarray}
with
\begin{equation*}
 \hat{V}_0(x) = (1-x)^2 \hat{v}_{11}(0) + 2x(1-x) \hat{v}_{12}(0) + x^2\hat{v}_{22}(0).
\end{equation*}

\subsection{Phase Separation at Constant Pressure}

Phase separation at constant pressure is possible provided that condition (\ref{eq:con3}) is violated,
which is only possible if $\Delta > 0$. Using $\hat{v}_{ij}(k=0)\!=\!\epsilon_{ij}^* R_{ij}^2 \pi$, we obtain
\begin{equation*}
 \Delta = \pi^2 \bigl[(\epsilon_{12}^*)^2 R_{12}^4 - \epsilon_{11}^* \epsilon_{22}^* R_{11}^2 R_{22}^2 \bigr] > 0.
\end{equation*}
The instable region of the phase diagram is also given by condition (\ref{eq:con3})
and we obtain the spinodal line from
\begin{equation*}
 1 + \rho \hat{V}_1(x) - \rho^2 x(1-x) \Delta = 0.
\end{equation*}
solving for the density leads to
\begin{equation}
\label{eq:spinodal_rho}
 \rho_s(x) = \frac{\hat{V}_1(x) + \sqrt{\hat{V}_1(x)^2 + 4x(1-x)\Delta}}{2x(1-x)\Delta}.
\end{equation}

To determine the bindodal line it is convenient to work with the Gibbs free energy density $g(x,p)$, where the pressure $p$ is the independent variable.
Thus we have to perform a Legendre transform of the free energy density $f(x,v)$
\begin{equation*}
 g(x,p) = f(x,v(p,x)) + p v(p,x) = f(x,\rho(p,x)) + \frac{p}{\rho(p,x)}.
\end{equation*}
Therefor we need the density $\rho$ as a function of pressure, which is obtained by inverting equation (\ref{eq:pressure}).
The quadratic equation in $\rho$ has two solution: one is negative and therefore physically irrelevant and the other one is given by

\begin{equation}
\label{eq:rho}
 \rho(p,x) = \frac{-1 + \sqrt{1+2 \beta p\hat{V}_0(x)}}{\hat{V}_0(x)}.
\end{equation}
Finally the Gibbs free energy density is
\begin{align*}
 \beta g(x,p) &= \ln(\lambda^2 \rho(p,x)) - 1 + (1-x)\ln(1-x) + x\ln(x) \\
 &+ \frac{1}{2}\rho(p,x) \hat{V}_0(x) + \frac{\beta p }{\rho(p,x)}.
\end{align*}
With this thermodynamic potential, the coexistence condition is given by
\begin{equation}
\label{eq:coexcondition_p}
 \biggl( \frac{\partial g}{\partial x} \biggr)_p \biggl|_{x_A} =   \biggl( \frac{\partial g}{\partial x} \biggr)_p \biggl|_{x_B} = \frac{g(x_A,p)-g(x_B,p)}{x_A-x_B}.
\end{equation}
Above a critical pressure $p_{crit}$ the derivative of the Gibbs free energy shows the typical loops. 
For fixed $p$ equation \ref{eq:coexcondition_p} can be solved numerically using the common tangent construction, which will not be presented here.
Resulting phase diagrams for various paramters of $R_{ij}$ and $\epsilon^*_{ij}$ can be found in reference \cite{Archer2001}

\section*{Appendix 2}\label{appendix2}

The gradient of a scalar field $f(\phi, \theta)$ defined on the surface of a sphere with radius $R$ is given by
\begin{equation}
 \nabla f(\phi,\theta) = \frac{1}{R^2 \sin \theta} \frac{\partial f}{\partial \phi} \mathbf{e_\phi} + \frac{1}{R} \frac{\partial f}{\partial \theta} \mathbf{e_\theta}.
\end{equation}
To numerically compute the gradient we use a central difference scheme for the $\phi$-component 
\begin{align}
 \bigl(\nabla f\bigr)_\phi (\phi_i,\theta_j) &= \frac{f(\phi_{i+1},\theta_j)-f(\phi_{i-1},\theta_j)}{2 R^2\, \sin\theta\,d\phi},
\end{align}
where $i=1, ... ,M-2$. 
We can easily extend the scheme over the edges of the numerical grid by using 
$(\phi_0$ and $\phi_{M-2})$ and $(\phi_1$ and $\phi_{M-1})$ to obtain the gradient at the position 
$(\phi_{M-1}$,$\theta_j)$ and at $(\phi_0,\theta_j)$. 
Similarly the $\theta$-component is computed via
\begin{align}
 \bigl(\nabla f \bigr)_{\theta}(\phi_i,\theta_j) &= \frac{f(\phi_i,\theta_{j+1}) - f(\phi_i,\theta_{j-1})}{2R\, d\theta}
\end{align}
where $j=1, ... ,N-2$. 
When computing this component of the gradient on the edges of the numerical grid, 
one has to keep in mind that the $j=0$ row of the array bends around the north pole
(also, the $j=N-1$ row bends around the south pole). 
The gradient of the points surrounding the north pole is thus given by
\begin{eqnarray}
 \bigl( \nabla f\bigr)_\theta (\phi_i,\theta_0) &=& \frac{f(\phi_i,\theta_1)-f(\phi_{i+M/2},\theta_0)}{2R\,d\theta},
\end{eqnarray}
where $i=0,\ldots,M/2-1$, and
\begin{eqnarray}
\bigl( \nabla f\bigr)_\theta (\phi_i,\theta_0) &=& \frac{f(\phi_i,\theta_1)-f(\phi_{i-M/2},\theta_0)}{2R\,d\theta},  
\end{eqnarray}
where $i=M/2,\ldots,M-1$.
We compute the gradient's $\theta$-component on points surrounding the south pole by using $\theta_{N-1}$ 
and $\theta_{N-2}$ on the r.h.s. 

The divergence of a vector field $\mathbf{A}(\phi,\theta)$ defined on the surface 
of a sphere is given by
\begin{equation}
\label{eq:divergence}
 \nabla \cdot\mathbf{A}(\phi,\theta) = \frac{1}{R \sin\theta}\frac{\partial}{\partial \phi} A_\phi + \frac{1}{R \sin\theta} \frac{\partial}{\partial \theta} (A_\theta \sin \theta).
\end{equation}
The finite difference method described above for the gradient can again be employed. 
However, when computing the second term on the r.h.s of \eqref{eq:divergence} it must be recalled 
that the $\theta$-component of a vector field on the sphere points in direction of the south pole.
On the edges ($\theta_{0}$ and $\theta_{N-1}$) this leads to a sign change in the finite difference scheme.
For points around the north pole the second term of equation (\ref{eq:divergence}) is given by
\begin{align}
 &\frac{\partial}{\partial \theta} \bigl( A_\theta(\phi_i,\theta_0) \sin \theta_0 \bigr) \\
 &= \frac{A_\theta(\phi_i,\theta_1)\sin \theta_1 + A_\theta(\phi_{i+M/2},\theta_0)\sin\theta_0}{2d\theta} 
\end{align}
where $i=0,\ldots, M/2-1$ and 
\begin{align}
 &\frac{\partial}{\partial \theta} \bigl( A_\theta(\phi_i,\theta_0) \sin \theta_0 \bigr) \\
 &= \frac{A_\theta(\phi_i,\theta_1)\sin \theta_1 + A_\theta(\phi_{i-M/2},\theta_0)\sin\theta_0}{2d\theta} 
\end{align}
where $i=M/2,\ldots, M-1$. 
For the points surrounding the south pole
\begin{align}
 &\frac{\partial}{\partial \theta} \bigl( A_\theta(\phi_i,\theta_{N-1}) \sin \theta_0 \bigr) \\
 &= \frac{-A_\theta(\phi_{i+M/2},\theta_{N-1})\sin \theta_{N-1} - A_\theta(\phi_{i},\theta_{N-2})\sin\theta_{N-2}}{2d\theta} 
\end{align}
where $i=0,\ldots, M/2-1$ and
\begin{align}
 &\frac{\partial}{\partial \theta} \bigl( A_\theta(\phi_i,\theta_{N-1}) \sin \theta_0 \bigr) \\
 &= \frac{-A_\theta(\phi_{i+M/2},\theta_{N-1})\sin \theta_{N-1} - 
A_\theta(\phi_{i},\theta_{N-2})\sin\theta_{N-2}}{2d\theta} 
\end{align}
where $i=M/2,\ldots, M-1$.

\section*{Appendix 3}
\label{app:appendix3}

To compute the interaction energy between two meso-spheres in various configurations the density field on the sphere needs to be rotated.
Because of the non-uniform spherical grid this is a non trivial task.
For rotations in $\phi$-direction one can simply map each point onto the neighboring point.
Unfortunately for rotations in direction of $\theta$ this is not possible (see also figure \ref{fig:numerical_grid}). 
Fortunately, we can work around this problem by performing the rotations in the space of spherical 
harmonic functions. One can compute rotation matrices, which act upon the coefficients of the 
spherical harmonic expansion and hence the rotation is done independently of the numerical grid.
In this appendix we show how to compute these rotation matrices.

An arbitrary rotation of a rigid body can be specified using the three Euler angles $\alpha$, $\beta$, $\gamma$.
In a Cartesian coordinate system, this rotation is generally defined as a rotation around the $z$-axis by angle $\alpha$, followed by a rotation around the new $y$-axis with angle $\beta$ and finally a rotation around the new $z$-axis with angle $\gamma$.
In  our spherical symmetric case any orientation of the density field can be achieved by using only angles $\alpha \in [0,2\pi)$ and $\beta \in [0,\pi]$. 
The rotated expansion coefficients $\hat{a}_l^{m'}$ can be expressed using the following rotation matrix
\begin{eqnarray*}
\hat{a}_l^{m'} &=& \sum_{m=-l}^{l} T_l^{m',m}(\alpha, \beta, \gamma) a_l^m \\
T_l^{m',m} &=& e^{-im\gamma} H_l^{m',m}(\beta) e^{im\alpha}.
\end{eqnarray*}
We see that the rotation in $\phi$-direction around the $x$-axis by angle $\alpha$ and $\beta$ is achieved by a 
simple multiplication with an exponential (the rotation matrix is diagonal).
Rotation in $\theta$-direction by angle $\beta$ is given through the matrix $H_l^{m',m}(\beta)$, which becomes larger as one goes to higher $l$-subspaces.
Here we show how to compute this matrix using recursion, slightly modified from that described in 
Ref.~\cite{Gumerov2014}.

{\bf Step 1}. We compute all $H_l^{0,m}(\beta)$ for $m = 0,...,l$ for every subspace 
$l$ up to $L+1$, where 
$L$ is the desired upper limit.
These coefficients are given by the associated Legendre polynomials $P_l^m(x)$
\begin{equation*}
 H_l^{0,m} = (-1)^m \sqrt{\frac{(n-|m|)!}{(n+|m|)!}} P_l^{|m|}(\cos(\beta)).
\end{equation*}
Using the symmetry rule
\begin{equation}
\label{eq:sym1}
 H_l^{m',m}(\beta) = H_l^{-m',-m}(\beta),
\end{equation}
we also obtain $H_l^{0,m}$ for $m=-1,...,-l$. 
Furthermore we use a second symmetry relation to get $H_l^{m',0}(\beta)$ for $-l\le m'\le l$
\begin{equation}
 \label{eq:sym2}
 H_l^{m',m}(\beta) = H_l^{m,m'}(\beta).
\end{equation}

{\bf Step 2}. In every subspace $l$, we compute $H_l^{1,m}(\beta)$ for $m=1,...,l$ using 
the following recursion
\begin{eqnarray*}
 H_l^{1,m}(\beta) &=& \frac{1}{b_{l+1}^0} \biggl \{ b_{l+1}^{-m-1} \frac{1-\cos(\beta)}{2} H_{l+1}^{0,m+1} \\
 &-& b_{l+1}^{m-1} \frac{1+\cos(\beta)}{2} H_{l+1}^{0,m-1} \\
 &-& a_l^m \sin(\beta) H_{l+1}^{0,m}\biggr \},
\end{eqnarray*}
with
\begin{equation*}
 a_l^m = \sqrt{\frac{(l+1+m)(l+1-m)}{(2l+1)(2l+3)}}
\end{equation*}
and
\begin{equation*}
 b_l^m = \text{sgn}(m) \sqrt{\frac{(l-m-1)(l-m)}{(2l+1)(2l+1)}}
\end{equation*}
\begin{equation*}
 \text{sgn}(m) = \begin{cases}
                  1 & m \ge 0 \\
                  -1 & m < 0.
                 \end{cases}
\end{equation*}

Using the symmetry relations (\ref{eq:sym1}) and (\ref{eq:sym2}) we also obtain $H_l^{-1,m}(\beta)$ for $m=-1,...,-l$, $H_l^{m',1}(\beta)$ for $m'=0,...,l$ and $H_l^{m',-1}(\beta)$ for $m'=-1,...,-l$.

{\bf Step 3}.  
We compute $H_l^{m'+1,m}(\beta)$ for $m'=1,...l-1$ and $m = m',...,l$ within every subspace $l$ using

\begin{eqnarray*}
H_l^{m'+1,m}(\beta) &=& \frac{1}{d_l^{m'}} \Bigl\{d_l^{m'-1} H_l^{m'-1,m}(\beta) \\ 
&-& d_l^{m-1} H_l^{m',m-1}(\beta) + d_l^m H_l^{m',m+1}(\beta) \Bigr\},
\end{eqnarray*}
where
\begin{equation*}
 d^m_l = \frac{\text{sgn}(m)}{2}\bigl[(l-m)(l+m+1)\bigr]^{1/2}.
\end{equation*}

With symmetry relation (\ref{eq:sym2}) we can complete all missing entries in the positive 
$m,m'$ - triangle in each subspace $l$. The negative $m,m'$-triangle is given though symmetry 
rule (\ref{eq:sym1}).

{\bf Step 4}. We compute $H_{l}^{-1,m}(\beta)$ in every subspace $l$ for $m=1,...,l$.
\begin{eqnarray*}
 H_l^{-1,m}(\beta) &=& \frac{1}{b_{l+1}^{0}} \biggl\{b_{l+1}^{m+1} \frac{1-\cos(\beta)}{2} H_{l+1}^{0,-m-1}(\beta) \\
 &-& b_{l+1}^{-m+1} \frac{1+\cos(\beta)}{2} H_{l+1}^{0,-m+1}(\beta) \\ 
 &-& a_l^{-m} \sin(\beta) H_{l+1}^{0,-m}(\beta) \biggr\}
\end{eqnarray*}
Again using symmetry relation (\ref{eq:sym1}) we can add the obtained values to the 
positive $m'$, negative $m$ - triangle.

{\bf Step 5}. Finally we compute the coefficients $H_l^{m'-1,m}(\beta)$ for $m'=-1,...,-l+1$ and $m = -m',...,l$
\begin{eqnarray*}
 H_l^{m'-1,m}(\beta) &=& \frac{1}{d_l^{m'-1}} \Bigl \{ d_l^{m'} H_l^{m'+1,m}(\beta)\\
 &+& d_l^{m-1} H_l^{m',m-1}(\beta) - d_l^m H_l^{m',m+1} \Bigr \},
\end{eqnarray*}
and complete the missing entries for the negative $m'$, positive $m$ triangle using symmetry rule 
(\ref{eq:sym2}).
The matrix entries of this triangle can then be projected onto the positive $m'$, negative $m$ - 
triangle with symmetry relation (\ref{eq:sym1}), which leaves us with the completed rotation matrix.


\begin{thebibliography}{}

\bibitem{onuki}
A.~Onuki, {\it Phase transition dynamics} 
(Cambridge University Press, Cambridge, 2007).

\bibitem{lenne2009}
P.-F.~Lenne and A.~Nicolas, Soft Matter {\bf 5} 2841 (2009).

\bibitem{lingwood2010}
D.~Lingwood and K.~Simons, Science {\bf 46} 327 (2010).

\bibitem{vink1}
R.L.C.~Vink, Soft Matter {\bf 5} 4388 (2009).

\bibitem{vink2}
T.~Fischer and R.L.C.~Vink, J.Chem.Phys. {\bf 134} 055106 (2011).

\bibitem{Li1994}
W. Li and J.C. Lee, Physica A, {\bf 202} 165-174 (1994)

\bibitem{Lee1994}
J.C. Lee, Physica A, {\bf 210} 127-138 (1994)

\bibitem{Ghosh2012}
A. Ghosh, J. Samuel, and S. Sinha, EPL (Europhysics Letters) {\bf 98} 30003 (2012).

\bibitem{Marenduzzo2013}
D. Marenduzzo and E. Orlandini, Soft Matter {\bf 9} 1178–1187 (2013).

\bibitem{Fischer2010}
T~Fischer and R~L~C Vink, J. Phys.: Condens. Matter {\bf 22} 104123 (2010).

\bibitem{bianchi}
E.~Bianchi,  J.~Largo,  P.~Tartaglia,  E.~Zaccarelli  and F.~Sciortino, 
Phys.Rev.Lett. {\bf 97} 168301 (2006).

\bibitem{likos_review}
E.~Bianchi, R.~Blaak and C.N.~Likos, Phys.Chem.Chem.Phys. {\bf 13} 6397 (2011).

\bibitem{glotzer}
S.C.~Glotzer and M.J.~Solomon,
Nat. Mater. {\bf 6} 557 (2007).

\bibitem{Stillinger1976}
F.~H. Stillinger, The Journal of Chemical Physics {\bf 65} 3968 (1976).

\bibitem{Archer2001}
A.~J. Archer and R.~Evans, Phys. Rev. E {\bf 64} (2001).

\bibitem{Archer2002}
A~J Archer and R~Evans, J. Phys.: Condens. Matter {\bf 14} 1131 (2002).

\bibitem{Archer2003}
A.~J. Archer and R.~Evans, The Journal of Chemical Physics {\bf118 }, 9726 (2003).

\bibitem{Archer2004}
A~J Archer, C~N Likos, and R~Evans, J. Phys.: Condens. Matter {\bf 16} L297 (2004).

\bibitem{Archer2005}
A.~J. Archer, R.~Evans, R.~Roth, and M.~Oettel, The Journal of Chemical Physics {\bf 122} 084513 (2005).s

\bibitem{Archer2006}
A.~J. Archer, M.~Schmidt, and R.~Evans, Phys. Rev. E {\bf 73} (2006).

\bibitem{Louis2000}
A.~A. Louis, P.~G. Bolhuis, and J.~P. Hansen, Phys. Rev. E {\bf 62} 7961–7972 (2000).


\bibitem{gardiner}
C.W. Gardiner, {\it Handbook of Stochastic Methods}, (Springer, Berlin (1985)).

\bibitem{dhont_book}
J.~K.~G. Dhont, {\it An introduction to dynamics of colloids} (Elsevier, Amsterdam, 1996).

\bibitem{marconi1999}
U.M.B.~Marconi and P.~Tarazona, J.Chem.Phys. {\bf 110} 8032 (1999). 
  

\bibitem{reinhardtbrader}
J.~Reinhardt and J.M.~Brader, Phys.Rev.E {\bf 85} 011404 (2012). 

\bibitem{Driscoll1994}
J.~R. Driscoll and D.~M. Healy, Advances in Applied Mathematics {\bf 15} 202–250 (1994).

\bibitem{ccSHT}
Christopher Cantalupo, ccsht

\bibitem{Frigo2005}
M.~Frigo and S.G. Johnson, Proceedings of the IEEE {\bf 93} 216–231 (2005).





\bibitem{Abraham1976}
F.~F. Abraham, J. Chem. Phys. {\bf 64}, 2660 (1976).

\bibitem{Bolhuis2001}
P.~G. Bolhuis, A.~A. Louis, J.~P. Hansen, and E.~J. Meijer, J. Chem. Phys {\bf 114} 4296 (2001).

\bibitem{Dautenhahn1994}
J. Dautenhahn and C.~K. Hall, Macromolecules {\bf 27} 5399–5412 (1994).

\bibitem{Dhont1996}
J.~K.~G. Dhont, J. Chem. Phys. {\bf 105} 5112 (1996).


\bibitem{Evans1979a}
R.~Evans and M.M. Telo~da Gama, Molecular Physics {\bf 38} 687–698 (1979).

\bibitem{Archer2004a}
A.~J. Archer and R.~Evans, The Journal of Chemical Physics {\bf 121} 4246 (2004).

\bibitem{Fischer2011}
T.~Fischer and R.~L.~C. Vink, The Journal of Chemical Physics {\bf 134} 055106 (2011).

\bibitem{Flory1950}
P.~J. Flory and W.~R. Krigbaum, J. Chem. Phys. {\bf 18} 1086 (1950).


\bibitem{Likos2001}
Christos~N. Likos, Physics Reports {\bf 348} 267–439 (2001).



\bibitem{Louis2000a}
A.~A. Louis, P.~G. Bolhuis, J.~P. Hansen, and E.~J. Meijer, Physical Review Letters {\bf 85} 2522–2525 (2000).

\bibitem{Gumerov2014}
N.~A. Gumerov and R. Duraiswami {\it Recursive computation of spherical harmonic rotation coefficients of large degree} 

\bibitem{Zhang2004}
Z. Zhang and S. C. Glotzer, Nano Lett. {\bf 4} 1407–1413 (2004)

\bibitem{Jackson2004}
A. M. Jackson, J. W. Myerson and F. Stellacci, Nature Materials, {\bf 3} 330--336 (2004)

\bibitem{Singh2007}
C. Singh, P. K. Ghorai, M. A. Horsch, A. M. Jackson, R. G. Larson, F. Stellacci and S. C. Glotzer, Phys. Rev. Lett., {\bf 99} 226106 (2007)

\end{thebibliography}
\end{document}